\documentclass[iop,apj,twocolumn,numberedappendix]{emulateapj}
\usepackage{hyperref,latexsym,graphicx,rotating,amsmath, epsfig, natbib}
\usepackage[usenames, dvipsnames]{color}

\newcommand\Beq{\begin{eqnarray}} 
\newcommand\Eeq{\end{eqnarray}}

\newcommand{\nn}{\nonumber}

\newcommand{\pomega}{\varpi}

\renewcommand{\vec}[1]{\boldsymbol{#1}}

\renewcommand{\dot}{\vec{\cdot}}

\newcommand{\grad}{\vec{\nabla}}
\renewcommand{\div}{ \grad \dot }

\renewcommand{\perp}{\!\bot}

\shorttitle{}
\shortauthors{Conduction in Low Ma Flows} 

\begin{document}

\title{Conduction in Low Mach Number Flows: Part I Linear \& Weakly Nonlinear Regimes \bigskip}
\author{Daniel Lecoanet}
\affil{Dept.\ Astronomy \& Theoretical Astrophysics Center, University of California, Berkeley, CA 94720}
\affil{Kavli Institute for Theoretical Physics, University of California, Santa Barbara, CA 93106}
\email{dlecoanet@berkeley.edu}
\author{Benjamin P.\ Brown}
\affil{LASP and Dept. Astrophysical \& Planetary Sciences,
University of Colorado, Boulder, CO 80309}
\affil{Kavli Institute for Theoretical Physics, University of California, Santa Barbara, CA 93106}
\author{Ellen G.\ Zweibel}
\affil{Dept.\ Astronomy, University of Wisconsin, Madison, WI 53706-1582}
\affil{Center for Magnetic Self Organization in Laboratory and Astrophysical Plasmas, University of Wisconsin, Madison, WI 53706, USA}
\affil{Kavli Institute for Theoretical Physics, University of California, Santa Barbara, CA 93106}
\author{Keaton J.\ Burns}
\affil{Dept.\ Physics, Massachusetts Institute of Technology, Cambridge, MA 02139}
\affil{Kavli Institute for Theoretical Physics, University of California, Santa Barbara, CA 93106}
\author{Jeffrey S.\ Oishi}
\affil{Dept.\ Physics, Farmingdale State College, Farmingdale, NY 11735}
\affil{Dept.\ Astrophysics, American Museum of Natural History, New York, NY 10024}
\affil{Kavli Institute for Theoretical Physics, University of California, Santa Barbara, CA 93106}
\author{Geoffrey M.\ Vasil}
\affil{School of Mathematics \& Statistics, University of Sydney NSW 2006, Australia}

\begin{abstract}
Thermal conduction is an important energy transfer and damping mechanism in astrophysical flows.  Fourier's law---the heat flux is proportional to the negative temperature gradient, leading to temperature diffusion---is a well-known empirical model of thermal conduction.  However, entropy diffusion has emerged as an alternative thermal conduction model, despite not ensuring the monotonicity of entropy.  This paper investigates the differences between temperature and entropy diffusion for both linear internal gravity waves and weakly nonlinear convection.  In addition to simulating the two thermal conduction models with the fully compressible Navier--Stokes equations, we also study their effects in the reduced, ``sound-proof'' anelastic and pseudo-incompressible equations.  We find that in the linear and weakly nonlinear regime, temperature and entropy diffusion give quantitatively similar results, although there are some larger errors in the pseudo-incompressible equations with temperature diffusion due to inaccuracies in the equation of state.  Extrapolating our weakly nonlinear results, we speculate that differences between temperature and entropy diffusion might become more important for strongly turbulent convection.
\end{abstract}
\slugcomment{}


\section{Introduction}\label{sec:Introduction}

In astrophysical fluid dynamics, important processes routinely occur on very disparate length and time scales.  Often, systems are driven on length scales orders of magnitude larger than the dissipation length scale.  Astrophysicists have turned to numerical simulations to attempt to gain insight into these complicated, nonlinear systems.  The inability to simulate the full range of spatial and temporal scales of a system have led to an ever-growing set of approximations, each of which has its own advantages and disadvantages.

For instance, the Navier--Stokes equations admit fast sound waves, which place strong restrictions on the time step of low Mach number flow when using an explicit time-integration scheme.  However, by removing the sound waves from the system of fluid equations, a ``sound-proof'' set of equations need not resolve the fast sound time scale.  These approximations range from the Boussinesq approximation, which assumes a constant density fluid; to the anelastic equations \citep{batchelor53,op62}, which assume small thermodynamic perturbations about a background state; to the pseudo-incompressible equations \citep{durran89,almgren06}, which allow for order unity thermodynamic perturbations in all quantities except the pressure.  In \citet[][hereafter B12]{brown12} \& \citet[][hereafter V13]{vasil13}, we show that certain ideal formulations of the anelastic \citep[e.g.,][]{lantz92,br95} and pseudo-incompressible \citep[e.g.,][]{durran89} equations reproduce internal gravity wave eigenfunctions and frequencies better than other formulations.

In this paper, we turn to non-ideal behavior in these different equation sets, focusing on thermal conduction.  Thermal conduction plays a role in damping internal gravity waves in the radiative zones of stars, and in setting the ferocity of convection (presumably related to the Rayleigh number, the ratio of driving to damping on large scales) in the convection zone of stars.  However, thermal conduction in convection is most important on length scales much smaller than the driving scale.  Thus, simulations either replace thermal conduction by algorithmic numerical conduction \citep[e.g., the Athena code,][]{gs08,stone08}, or an explicit conduction term which acts on much larger length scales than in the physical system \citep[e.g.,][]{clune99,brun04,nonaka10}.  Simulations with a reduced dynamic range are feasible with current computational resources.

When increasing the strength of thermal conduction (and similarly viscosity) to remove small scales from the system, one has to decide how to model the neglected small scales.  One perspective is to run a direct numerical simulation (DNS), in which one uses the real damping processes that act on small scales.  In this case, thermal conduction is modeled by Fourier's law of conduction \citep{fourier22},
\Beq
\vec{Q}_T = -\kappa_T \grad T,
\Eeq
where $\vec{Q}$ is the heat flux, $\kappa_T$ is the conductivity, and $T$ is the temperature.  This leads to temperature diffusion.  However, it is computationally infeasible to use the microscopic diffusivities of many physical systems, so the diffusivities must be artificially increased.  Thus, even a DNS employs a certain sub grid-scale (SGS) model of thermal conduction.

Another perspective is to use a SGS model to describe how the unresolved small scales influence thermal conduction.  In this case, the unresolved {\it convective} heat flux can represented by a {\it conductive} heat flux.  Although there are many SGS models \citep[e.g.,][]{lesieur90}, one particularly popular model is the entropy diffusion model (used extensively in anelastic simulations, e.g., \citealt{clune99}, and also occasionally in fully compressible simulations, \citealt{cs86,cs89}),
\Beq
\vec{Q}_S = -\kappa_S \grad S,
\Eeq
where $S$ is the entropy.  In this paper we compare the temperature diffusion and entropy diffusion models.  \citet{glatzmaier84} argues that the heat liberated by an eddy is given by the local entropy gradient.  \citet{br95} argue for the ``engineering approach'' that the flux of entropy should be linear in the entropy gradient (though not necessarily parallel to it).  Practically speaking, perhaps the most important feature of entropy diffusion is that it does not require the calculation of the pressure perturbation, which can be advantageous for the anelastic equations.

\citet{calkins14} has recently calculated the onset of convection in anelastic simulations with either temperature diffusion or entropy diffusion, as well as in fully compressible Navier--Stokes simulations with temperature diffusion.  They find that the anelastic and Navier--Stokes equations with temperature diffusion have almost identical behavior, provided the background entropy gradient is close to adiabatic.  However, they only find ``qualitative'' rather than ``quantitative'' agreement between the temperature diffusion and entropy diffusion models.

The remainder of the paper is structured as follows.  First, in section~\ref{sec:Motivation}, we show that entropy diffusion can lead to non-monotonicity of entropy.  We state the various equations we use in section~\ref{sec:Model Equations} (and describe their numerical implementation in appendix~\ref{sec:dedalus equations}).  Next we study thermal damping of internal gravity waves, both numerically (section~\ref{sec:IGW numerics}) and analytically (section~\ref{sec:IGW analytics}).  In section~\ref{sec:steady convection} we describe convective steady states for each of our equation sets with either temperature or entropy diffusion.  Finally, section~\ref{sec:conclusion} summarizes our results, discusses its connection with other work, and suggests future paths of inquiry.

\section{Preliminaries}

\subsection{Motivation}\label{sec:Motivation}

A fundamental law of equilibrium statistical mechanics is that the total entropy of a closed system increases monotonically with time.  This is encapsulated in the entropy equation for a fully compressible fluid,
\Beq\label{entropy temp}
\rho T\frac{dS}{dt}=\div \kappa_T \grad T,
\Eeq
where $\rho, T, S$ are the density, temperature, and specific entropy of the fluid, and $d/dt=\partial_t+\vec{u}\dot\grad$ denotes the material derivative where $\vec{u}$ is the fluid velocity.  We will assume that the heat flux is proportional to $\grad X$, for $X=S, T$, and take $\kappa_X$ to be the constant of proportionality: $\vec{Q}=-\kappa_X\grad X$.  For temperature diffusion ($X=T$), $\kappa_T$ is the conductivity.  Using the continuity equation, this can be rewritten as
\Beq\label{entropy per volume temp}
\frac{\partial\rho S}{\partial t}=-\div\left[\rho S\vec{u} -\frac{\kappa_T}{T}\grad T\right]+\kappa_T\frac{|\grad T|^2}{T^2}.
\Eeq
Assuming boundary conditions that ensure that the term in the total divergence on the RHS of equation~\ref{entropy per volume temp} is zero on the boundaries (e.g., no penetration and no heat flux), the volume integral of equation~\ref{entropy per volume temp} shows entropy increases monotonically with time,
\Beq\label{second law}
\frac{\partial}{\partial t} \int_V \rho S \ dV = \int_V \kappa_T\frac{|\grad T|^2}{T^2} \ dV \geq 0.
\Eeq

However, if we instead use entropy diffusion in equation~\ref{entropy temp},
\Beq\label{entropy both}
\rho T\frac{dS}{dt}= \div \kappa_S \grad S,
\Eeq
the entropy per volume instead evolves according to
\Beq\label{entropy per volume both}
\frac{\partial\rho S}{\partial t}=-\div\left[\rho S\vec{u} - \frac{\kappa_S}{T}\grad S\right]  + \kappa_S\frac{\grad S \dot \grad T}{T^2}.
\Eeq

Again assuming boundary conditions such that the term in the total divergence is zero on the boundaries, the volume integral of equation~\ref{entropy per volume both} is
\begin{align}\label{modified second law}
\frac{\partial}{\partial t} \int_V \rho S dV =  \int_V \kappa_S\frac{\grad S \dot \grad T}{T^2} \ dV.
\end{align}
The $\grad S\dot\grad T$ term is not positive definite, so there is no guarantee that entropy increases monotonically with time.  Although we focus only on thermal diffusion and entropy diffusion, the only heat flux which will monotonically increase entropy is proportional to $\grad T$ \citep{ll59}\footnote{The most general heat flux which monotonically increases entropy is $Q_i\sim - M_{ij} \partial T/\partial x_j$, where $M_{ij}$ is a symmetric rank-2 tensor.  For instance, in relatively collisionless plasmas, the heat flux is carried by electrons which follow magnetic field lines.  Thus, the heat flux is in the direction of the local magnetic field, $\vec{Q}\sim -\vec{b}\vec{b}\dot\grad T$, where $\vec{b}$ is the unit vector in the direction of the magnetic field \citep[e.g.,][]{spitzer62,balbus00}.  In this case, $M_{ij} = b_i b_j$ is symmetric, so entropy will increase monotonically.}.

This paper investigates the effects of this modification of the second law of thermodynamics for linear waves and weakly nonlinear convective equilibria.

\subsection{Model Equations}\label{sec:Model Equations}

We break the thermodynamic variables into background and fluctuating parts, e.g., $S = \overline{S} + S'$.  The background fields are time independent and satisfy hydrostatic and thermal equilibrium,
\Beq\label{hydrostatic equilibrium}
\grad \overline{P} = \vec{g}\overline{\rho}, \\
\grad \dot \overline{\vec{Q}} = 0, \label{thermal equilibrium}
\Eeq
where $P$ is the pressure, $\vec{g}$ is the gravitational acceleration, and $\vec{Q}$ is the heat flux.  This paper studies the effects of varying the form of $\vec{Q}$.  To simplify the problem, we assume the fluid to be an ideal gas with constant ratio of specific heats $\gamma$.  We make extensive use of the linearized thermodynamic relations,
\Beq\label{linearized thermo 1}
\frac{P'}{\overline{P}} & = & \frac{\rho'}{\overline{\rho}} + \frac{T'}{\overline{T}}, \\
\frac{S'}{C_P} & = & \frac{P'}{\gamma \overline{P}} - \frac{\rho'}{\overline{\rho}}. \label{linearized thermo 2}
\Eeq

\subsubsection{Full Compressible Equations}\label{sec:FC}

The fully compressible (FC) equations are
\Beq
&&\rho\left(\partial_t\vec{u}+\vec{u}\dot\grad\vec{u}\right) + \grad P = \vec{g}\rho - \grad\dot\Pi, \label{FC momentum nonlinear} \\
&& \partial_t\rho + \vec{u}\dot\grad\rho + \rho \grad\dot\vec{u}=0,  \label{FC continuity nonlinear} \\
&& \partial_t S + \vec{u}\dot\grad{S} = - \frac{1}{\rho T}\grad\dot\vec{Q} - \frac{1}{\rho T}\Pi_{ij}\partial_{x_i}u_j, \label{FC entropy nonlinear}
\Eeq
where repeated indices are summed over, $\Pi_{ij}$ is the viscous stress tensor,
\Beq
\Pi_{ij} = -\mu \left(\partial_{x_i}u_j + \partial_{x_j}u_i - \frac{2}{3}\delta_{ij}\grad\dot\vec{u}\right),
\Eeq
and $\delta_{ij}$ is the Kronecker delta.

To study waves, we solve the linearized, inviscid, FC equations.  We subtract off hydrostatic equilibrium and thermal equilibrium (equations~\ref{hydrostatic equilibrium} \& \ref{thermal equilibrium}), and linearize the thermodynamic variables (equations~\ref{linearized thermo 1} \& \ref{linearized thermo 2}).  To use notation consistent with B12, we pick $S'$ and $\varpi'=P'/\overline{\rho}$ as our thermodynamic variables.  Then the FC equations take the form
\Beq
&& \partial_t\vec{u} + \grad\varpi' - \varpi'\grad\left(\frac{\overline{S}}{C_P}\right) = -\vec{g} \frac{S'}{C_P}, \label{FC momentum} \\
&& \partial_t S' + \vec{u}\dot\grad \overline{S} = - \frac{1}{\overline{\rho} \overline{T}}\grad\dot\vec{Q}', \label{FC entropy} \\
&& \vec{u}\dot\grad\left(\frac{\overline{S}}{C_P}\right) + \vec{u}\dot\grad\log\overline{\rho} + \grad\dot\vec{u} = \nn \\
&& \quad \quad -\frac{1}{\overline{c_s}^2}\partial_t \varpi' - \frac{1}{\overline{\rho}\overline{T}C_P}\grad\dot\vec{Q}', \label{FC pressure}
\Eeq
where $\overline{c_s}^2=\gamma\overline{P}/\overline{\rho}$ is the adiabatic sound speed.  These equations support sound waves because they include the $\partial_t\varpi'/\overline{c_s}^2$ term in equation~\ref{FC pressure}.  Also note that thermal conduction appears in both thermodynamic equations.

\subsubsection{Pseudo-Incompressible Equations}\label{sec:PI}

The pseudo-incompressible (PI) equations (see V13) assume that sound waves rapidly equilibrate pressure fluctuations, so that pressure fluctuations are small ($\mathcal{O}(\overline{P}{\rm Ma}^2)$, where ${\rm Ma}$ is the Mach number) when averaged over a sound crossing time.  The pressure fluctuations must be retained in the pressure gradient term in the momentum equation to keep the flow from building large pressure fluctuations, but must be dropped everywhere else.  The PI equations are
\Beq
&&\rho\left(\partial_t\vec{u}+\vec{u}\dot\grad\vec{u}\right) + \overline{\beta}\grad \left(\frac{\pi'}{\overline{\beta}}\right) = \vec{g}\rho' - \grad\dot\Pi, \label{PI full momentum}\\
&& \partial_t\rho + \vec{u}\dot\grad\rho + \rho \grad\dot\vec{u}=0, \label{PI full continuity} \\
&& \vec{u}\dot\grad \overline{P} + \gamma \overline{P}\grad\dot\vec{u} =  \nonumber \\
&& \quad\quad\quad\quad\quad - \frac{1}{C_V}\grad\dot\vec{Q}  - \frac{1}{C_V} \Pi_{ij}\partial_{x_i}u_j, \label{PI full constraint}
\Eeq
where $\overline{\beta} = \overline{P}^{1/\gamma}$ and $C_V$ is the specific heat at constant volume.  In the equation of state, $P$ is replaced by $\overline{P}$, i.e., $T=T(\rho,\overline{P})$ and $s=s(\rho,\overline{P})$.  The variable $\pi'$ is the $\mathcal{O}(\overline{P}{\rm Ma}^2)$ correction to the background pressure.

The linearized, inviscid PI equations are very closely related to the linearized, inviscid FC equations: the $\varpi'$ term in equation~\ref{FC pressure} is dropped,
\Beq
&& \partial_t\vec{u}+ \grad\varpi' - \varpi'\grad\left(\frac{\overline{S}}{C_P}\right) = -\vec{g} \frac{S'}{C_P}, \label{PI momentum} \\
&& \partial_t S' + \vec{u}\dot\grad \overline{S} = - \frac{1}{\overline{\rho} \overline{T}}\grad\dot\vec{Q}', \label{PI entropy} \\
&& \vec{u}\dot\grad\left(\frac{\overline{S}}{C_P}\right) + \vec{u}\dot\grad\log\overline{\rho} + \grad\dot\vec{u} = \nn \\
&& \quad \quad \quad \quad \quad \quad \quad \quad \quad \quad \quad \quad   - \frac{1}{\overline{\rho}\overline{T}C_P}\grad\dot\vec{Q}', \label{PI pressure}
\Eeq
and the PI equations use the modified equation of state (compare to equation~\ref{linearized thermo 2}),
\Beq\label{PI equation of state}
\frac{S'}{C_P} =  - \frac{\rho'}{\overline{\rho}}.
\Eeq

\subsubsection{Anelastic Equations}\label{sec:ANS}

The anelastic (AN) equations (see B12) were first used in astrophysics to remove sound waves from convection simulations.  Efficient convection almost entirely erases an unstable entropy gradient.  Thus, the anelastic equations are derived in the limit that $\grad (\overline{S}/C_P) \sim \mathcal{O}({\rm Ma}^2/L_z) \ll 1$, where $L_z$ is the vertical (or radial) length of the convection zone.  Furthermore, the AN equations assume that all thermodynamic fluctuations are $\mathcal{O}({\rm Ma}^2)$, and thus the linearized thermodynamic relations (equations~\ref{linearized thermo 1} \& \ref{linearized thermo 2}) can be used.  The AN equations can be written
\Beq
&& \partial_t\vec{u}+\vec{u}\dot\grad\vec{u} + \grad\varpi'  = -\vec{g} \frac{S'}{C_P} - \grad\dot\Pi, \label{AN momentum} \\
&& \partial_t S' + \vec{u}\dot\grad{S'} + \vec{u}\dot\grad \overline{S} = \nonumber \\
&&\quad\quad\quad\quad\quad\quad\quad- \frac{1}{\overline{\rho} \overline{T}}\grad\dot\vec{Q}' - \frac{1}{\overline{\rho}\overline{T}} \Pi_{ij}\partial_{x_i}u_j, \label{AN entropy} \\
&&  \vec{u}\dot\grad\log\overline{\rho} + \grad\dot\vec{u} = 0.\label{AN pressure}
\Eeq
Having already linearized the thermodynamics, these equations bear striking similarity to the linearized FC \& PI equations, although they include the nonlinear $\vec{u}\dot\grad \vec{u}$ and $\vec{u}\dot\grad S'$ terms.  The $\grad \overline{S}/C_P$ terms in the momentum equation and constraint equation have been dropped, as well as the heating terms on the RHS of the constraint equation (which can be justified by dimensional analysis).

The linearized, inviscid AN equations are
\Beq\label{AN linear mom}
&& \partial_t\vec{u} + \grad\varpi'  = -\vec{g} \frac{S'}{C_P}, \\
&& \partial_t S'  + \vec{u}\dot\grad \overline{S} = - \frac{1}{\overline{\rho} \overline{T}}\grad\dot\vec{Q}' \label{AN linear entropy} \\
&&  \vec{u}\dot\grad\log\overline{\rho} + \grad\dot\vec{u} = 0. \label{AN linear constraint}
\Eeq

\section{Linear Wave Modes: Numerics}\label{sec:IGW numerics}

We solve for internal gravity wave (IGW) eigenmodes with different thermal conduction models using Dedalus\footnote{For more information and links to the source code, see \url
{dedalus-project.org}.} \citep{burns14}. Dedalus is a general framework for studying partial differential equations, including eigenvalue problems, boundary value problems, and initial value problems (i.e., simulations). It uses the $\tau$ spectral method to solve nearly arbitrary equation sets including algebraic constraints and complex boundary conditions. This flexibility allows us to specify the linear eigenvalue problem for IGWs in all three equation sets discussed above, with different thermal conduction models, all within the same code. In all cases, we use a 2D Cartesian domain with a Fourier grid in the horizontal ($x$) and a Chebyshev grid in the vertical ($z$) directions. In section~\ref{sec:steady convection}, we use Dedalus to evolve the nonlinear versions of these equation sets in time, and in appendix~\ref{sec:dedalus equations} we specify the exact equations as entered into the code.

We use a polytrope background field:
\begin{align}\label{polytrope T}
\overline{T} & = T_0 \frac{L_z+H-z}{H}, \\
\overline{\rho} & = \rho_0 \left(\frac{L_z+H-z}{H}\right)^n, \\
\overline{P} & = P_0 \left(\frac{L_z+H-z}{H}\right)^{n+1}, \label{polytrope P}
\end{align}
where $\rho_0, T_0, P_0$ are constants satisfying $\rho_0 T_0=P_0$, $L_z$ is the box height, and $H$ is the local scale height at the top of the box.  $n$ is the polytropic index, and satisfies $\vec{g} = - T_0(n+1)/H \vec{e}_z$.  We non-dimensionalize the system by setting $\rho_0=T_0=P_0=H=1$.  We take $\gamma=5/3$, so any $n>1.5$ corresponds to stable stratification---we pick $n=2$.  We take the box size to be $(L_x,L_z)=(78.3, 26.1)$, which corresponds to $\approx 6.6$ density scale heights, the number of density scale heights in the solar radiative zone.  The vertical resolution is typically 128 grid points \& modes (no dealiasing is needed for linear calculations).

This background satisfies thermal equilibrium only when using temperature diffusion with a constant $\overline{\kappa_T}$.  We assume that the background fields and the perturbation fields conduct heat differently.  Various authors \citep[e.g.,][]{br95,clune99,jones09} have argued that the heat conduction acting on the perturbation fields is actually a SGS effect from unresolved turbulent motions, and thus different from the microphysical heat conduction acting on the background fields.  In our calculations, the background fields conduct heat using temperature diffusion with a constant $\overline{\kappa_T}$; we use different choices for the conduction model for the perturbation fields.  The equations remain consistent as the perturbations never feed back onto the background fields.

The wave perturbations evolve according to the equations described in sections~\ref{sec:FC}--\ref{sec:ANS}.  We use two forms for $\vec{Q}'$,
\Beq\label{Q_T}
\vec{Q}_T' = - \chi_T \overline{\rho} \grad T',\\
\vec{Q}_S' = - \chi_S \overline{\rho}\overline{T}\grad \frac{S'}{C_P}, \label{Q_S}
\Eeq
where $\chi_T=\chi_S=10^{-5}$ are taken to be constant, which implies a constant diffusivity throughout the domain.  Such a small diffusivity ensures all modes are weakly damped.  We will refer to these two thermal conduction models as $T$-diffusion and $S$-diffusion.  Our boundary conditions are $w=0$ (the vertical velocity) and $Q_z=0$ at $z=0$ and $z=L_z$, and periodic in the horizontal direction.  The eigenmodes depend on horizontal and vertical wave numbers.  We define the $(n,m)$ mode to be the mode with horizontal wavenumber $k_x=2\pi n/L_x$, and $m$ extrema in the vertical direction (with vertical wavenumber defined as $k_z=2\pi m/L_z$).  $n$ and $m$ are the mode's horizontal and vertical mode number respectively.  The total wavenumber of a mode is $k=\sqrt{k_x^2+k_z^2}$.

\begin{figure}
  \begin{center}
    \includegraphics[width=9cm]{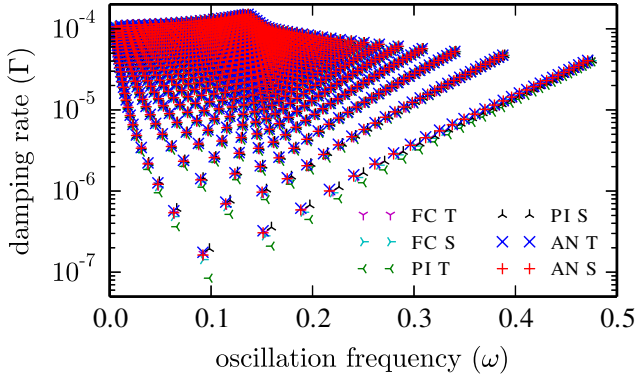}
  \end{center}
\caption{Damping rates and oscillation frequencies of gravity wave modes of the FC, PI, and AN equations, using either $T$- or $S$- diffusion.  The first thirty radial and horizontal modes are shown for each equation set.  The modes with lowest mode number are at the bottom of the plot.  Moving up and to the left corresponds to increasing the vertical mode number ($m$), and moving up and to the right corresponds to increasing the horizontal mode number ($n$).
\label{fig:eigenvalues}
}
\end{figure}

The linear eigenvalues vary in time as $\exp(-i\omega - \Gamma t)$, where $\omega$ is the oscillation frequency and $\Gamma$ is the damping rate. Figure~\ref{fig:eigenvalues} shows eigenvalues of the different equation sets with either $T$- or $S$- diffusion.  To ensure the accuracy of the eigenvalues, we compare the damping rate, $\Gamma$, to the analytic expression of the damping rate given in equations~\ref{AN theta_T} \& \ref{AN theta_S} and the analogous expressions in appendices~\ref{sec:FC analytics} \& \ref{sec:PI analytics}.  In all cases, the discrepancy is less than 1\%, and typically is less than $0.01\%$.

\begin{figure}
  \begin{center}
    \includegraphics[width=9cm]{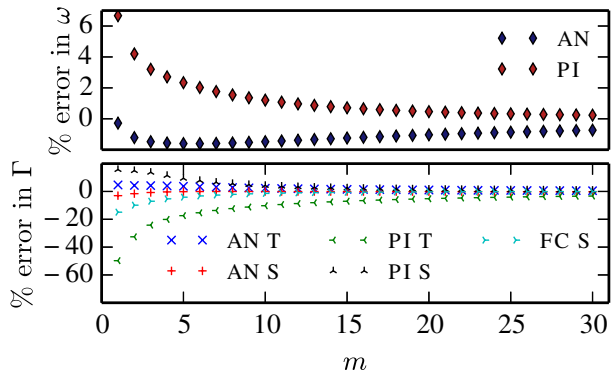}
  \end{center}
\caption{Percent error in oscillation frequencies and damping rates between different equations and thermal conduction models and the FC equations with $T$-diffusion, e.g., $(\omega_{AN;T}-\omega_{FC;T})/\omega_{FC;T}$.  The modes have $n=1$, but with varying vertical mode number $m$---these are the modes with the largest errors.  The percent error in oscillation frequency depends only on the equation set, not on the model of thermal conduction.
\label{fig:eigenvalue difference}
}
\end{figure}

The damping rates and oscillation frequencies match very well for the different equation sets and conduction models, particularly for large $k H$.  For small $k H$, there are some discrepancies in both damping rates and oscillation frequencies.  Figure~\ref{fig:eigenvalue difference} shows the percent error in the oscillation frequency and damping rate with respect to the eigenvalues of the FC equations with $T$-diffusion.  The relative errors are plotted for the modes $(1,m)$, with $m$ ranging from one to thirty.

The largest errors in damping rate occur for the PI equations with $T$-diffusion, where the damping rate is underestimated by half for the $(1,1)$ mode---interestingly, the PI equations with $S$-diffusion seems to agree more closely with the true damping rate.  This may be because the PI equations do not use the full linearized equation of state to calculate the temperature (compare equations~\ref{linearized thermo 1} \& \ref{linearized thermo 2} to equation~\ref{PI equation of state}).

For the AN equations, the error in damping rate is always less than twenty percent.  For all the models described here, the relative errors are less than 10\% for mode numbers greater than ten (or four if we neglect the PI equations with $T$-diffusion).  The FC equations with $S$-diffusion have relative errors of less than 1\% for mode numbers greater than eleven, whereas the AN equations with $S$-diffusion have relative errors of less than 1\% for mode numbers greater than nine (which is much better than the AN equations with $T$-diffusion).

Overall, there is little difference in the damping rates between $T$-diffusion and $S$-diffusion.  We will explain this by studying the linear problem analytically in section~\ref{sec:IGW analytics}.

In addition to errors in damping rates, there are also small (several percent) errors in oscillation frequency associated with using different equation sets.  Although the PI equations have the largest relative error (almost seven percent) for the $(1,1)$ mode, the AN equations have more persistent errors as the vertical mode number increases.

As shown in B12 \& V13, the ideal linear eigenvalues differ more among the different equation sets in spherical geometry than in plane parallel geometry.  Furthermore, in spherical geometry, the eigenfunctions also differ between equation sets.  These suggest that differences between damping rates (which depend on the eigenfunctions, as shown in section~\ref{sec:IGW analytics}) and oscillation frequencies will be larger in spherical geometry than in plane parallel geometry.  However, the differences in spherical geometry will also become small for $kH\gg 1$.

Nonlinear damping can also be an important damping mechanism, especially for modes with low linear damping rates.  Via nonlinear interactions, low wavenumber gravity waves can couple and transfer energy to higher wavenumber gravity waves \citep[e.g.,][]{weinberg12}.  The energy in high wavenumber modes can then be damped via dissipative effects, e.g., thermal conduction.  Thus, it is possible that the low wavenumber modes which have the largest discrepancies in linear damping rates could still be damped at the correct rate in fully nonlinear simulations---if the damping is dominated by nonlinearities.

\section{Linear Wave Modes:  Analytics}\label{sec:IGW analytics}

In the previous section, we demonstrated that IGWs have very similar damping rates with either $T$- or $S$- diffusion.  To better understand why this is the case, we study the linear IGW problem analytically.  We use different approximations to render the problem tractable.  First, we derive the eigenvalue equation in the large wavenumber limit.  Second, we assume dissipation is weak, and derive an expression for the damping rate.  The numerical results presented above in section~\ref{sec:IGW numerics} satisfy this weak dissipation assumption.  We only include the details of the calculations for the AN equations; the (similar) main results for the FC \& PI equations can be found in appendices~\ref{sec:FC analytics} \& \ref{sec:PI analytics}, respectively.  We find that $T$- and $S$- diffusion give the same damping rate because $T$ and $S$ are approximately proportional to one another in the large wavenumber limit.

\subsection{Large Wavenumber Limit}

We derive the eigenvalue equation for the AN equation (see appendices~\ref{sec:FC analytics} \& \ref{sec:PI analytics} for the eigenvalue equations for the FC and PI equations, respectively), for both $T$- and $S$- diffusion.  In the limit of $kH\gg 1$, these eigenvalue equations are equivalent, which implies that $T$- and $S$- diffusion will give the same eigenfunctions and eigenvalues.

To simplify the expressions, we will drop all terms with derivatives on background quantities, which are order $(k H)^{-1}\ll 1$.  In this limit, the eigenvalue equation for the AN equations using $T$-diffusion is
\Beq
&&\left[1+\frac{i\overline{\kappa_T}}{\overline{\rho}C_P\omega}\nabla^2 + \frac{i\overline{\kappa_T}}{\overline{\rho}C_P\omega}\frac{\gamma-1}{\gamma} \frac{g}{\overline{T}} \partial_z\right]\nabla^2w \nonumber \\
&& \quad \quad \quad \quad \quad \quad \quad \quad \quad \quad \quad \quad = -\frac{k_{\perp}^2\overline{N}^2}{\omega^2} w,
\Eeq
where $\overline{N}^2 = g \partial_z\overline{S}/C_P$ is the squared buoyancy (Brunt--V\"{a}is\"{a}l\"{a}) frequency.  The eigenvalue equation using $S$- diffusion is
\Beq
\left[1+\frac{i\overline{\kappa_S}}{\overline{\rho}\overline{T}\omega}\nabla^2\right] \nabla^2 w = -\frac{k_{\perp}^2\overline{N}^2}{\omega^2} w.
\Eeq
Note that $\nabla^2\gg (g/\overline{T})\partial_z$, because $g/\overline{T}\sim H^{-1}$.  Thus, the two eigenvalue equations are equivalent under the identification $\overline{\kappa_S}=\overline{\kappa_T}\overline{T}/C_P$.  This shows that $T$- and $S$- diffusion will have the same eigenfunctions and eigenvalues in the limit of large $kH$, as can be seen in figure~\ref{fig:eigenvalues}.
 
\subsection{Damping Rates \& the Weak Dissipation Limit}\label{sec:damping rates}

By manipulating the equations of motion, the damping rate can be expressed as a ratio of volume averages of the eigenfunctions.  In this section, we focus on the AN equations; the analogous results for FC \& PI equations are in appendices~\ref{sec:FC analytics} \& \ref{sec:PI analytics}, respectively.  Dotting the momentum equation (\ref{AN linear mom}) with $\overline{\rho}\vec{u}$ gives an energy equation,
\Beq
\partial_t\left(\frac{1}{2}\overline{\rho}|\vec{u}|^2\right) = \grad\dot\left(\overline{\rho}\vec{u}\varpi'\right) + \frac{g\overline{\rho}}{C_P} S' w.
\Eeq
The vertical velocity is given by the entropy equation (\ref{AN linear entropy}),
\Beq\label{vertical velocity}
w = -\frac{\partial_t S'}{\partial_z\overline{S}} - \frac{1}{\overline{\rho}\overline{T}\partial_z\overline{S}} \grad\dot\vec{Q}'.
\Eeq
Using this relation, we can rewrite the energy equation in the form
\Beq
\partial_t E + \grad\dot\vec{F} = - \theta,
\Eeq
where
\Beq
E & = & \frac{1}{2}\overline{\rho}|\vec{u}|^2 + \frac{1}{2}\frac{g\overline{\rho}}{C_P} \left(\partial_z\overline{S}\right)^{-1} S'^2, \\
\vec{F} & = & \overline{\rho} \vec{u}\varpi' + \frac{g}{\overline{T}\partial_z\overline{S}} \frac{S'}{C_P} \vec{Q}, \\
\theta & = & -\vec{Q}\dot\grad\left(\frac{g}{\overline{T}\partial_z\overline{S}} \frac{S'}{C_P}\right),
\Eeq
are the wave energy, the energy flux, and the change of wave energy due to thermal conduction, respectively.

The damping rate (of the perturbations) $\Gamma$ is
\Beq
\Gamma = \frac{\left\langle\theta\right\rangle}{2 \left\langle E\right\rangle},
\Eeq
where $\left\langle\cdot\right\rangle$ denotes a volume average.  For $T$- and $S$- diffusion, the expressions for $\theta$ are
\Beq\label{AN theta_T}
\theta_T & = & \overline{\kappa_T} \grad\left(\frac{g S'}{C_P\overline{T} \partial_z\overline{S}}\right) \dot\grad \left(\overline{T} \frac{S'}{C_P} + \frac{\gamma-1}{\gamma}\varpi' \right), \\
\theta_S & = & \overline{\kappa_S}\grad\left(\frac{g S'}{C_P\overline{T} \partial_z\overline{S}}\right)\dot\grad S'.\label{AN theta_S}
\Eeq
These expressions follow directly from the equations of motion and contain no approximations.  They are used to check the numeric damping rates calculated in figure~\ref{fig:eigenvalues}.

In equation~\ref{AN theta_T}, we have rewritten $T'$ as a function of $S'$ and $\varpi'$ using the linearized equation of state (equations~\ref{linearized thermo 1} \& \ref{linearized thermo 2}).  $T'$ is comprised of a part that is proportional to $S'$ and a part that is proportional to $\varpi'$.  We will show below that if $k H\gg 1$, the $\varpi'$ term is much smaller than the $S'$ term.  Thus, $T'$ and $S'$ are well aligned, and give the same damping rate.

If we assume $g/(\overline{T}\partial_z\overline{S})$ is constant (as is the case for a plane-parallel polytrope atmosphere), then $\theta_S$ is positive definite, proving there are no overstable modes with $S$-diffusion.  By contrast, $\theta_T$ cannot be shown to be positive definite, leaving open the possibility of overstable IGWs.  For the FC \& PI equations, neither $\theta_S$ nor $\theta_T$ can be shown to be positive definite (see appendices~\ref{sec:FC analytics} \& \ref{sec:PI analytics}).  \citet{lou90} searched extensively for overstable IGWs using the FC equations with temperature diffusion in a polytrope atmosphere, but found none.  We also have not found any overstable IGWs.

Next we calculate the relative sizes of the different terms in equations~\ref{AN theta_T} \& \ref{AN theta_S} by solving for the eigenfunctions.  To simplify the calculation, we will use the weak dissipation limit.  For sufficiently small dissipation, the eigenfunctions are very well approximated by the ideal eigenfunctions, which can be solved for analytically if we use the WKB approximation (assuming $kH\gg 1$).

The ideal eigenvalue equation (see, e.g., B12) is
\Beq
\omega^2\left(k_{\perp}^2 - \partial_z^2\right) w -&&\omega^2\partial_z\left((\partial_z\log\overline{\rho})w\right) \nn \\
&& \quad \quad \quad \quad \quad= \overline{N}^2k^2_{\perp} w,
\Eeq
The lowest order WKB approximation to the solution is
\Beq
w\approx \frac{A}{\sqrt{\overline{\rho}k_z}}\exp\left(-\int_{z_0}^z k_z(z') \ dz'\right),
\Eeq
where $A=w(z_0)\sqrt{\overline{\rho}k_z}$ is the amplitude and $k_z(z)$ satisfies
\Beq
k_z^2 = k_{\perp}^2\left(\frac{\overline{N}^2}{\omega^2}-1\right).
\Eeq
The entropy and pressure perturbations are related to one another via
\Beq
S' & = & -i\frac{w\partial_z\overline{S}}{\omega}, \\
\varpi' & = & \frac{i\omega}{k_{\perp}^2}\left(\partial_z w + w\partial_z\log\overline{\rho}\right) \approx \frac{\omega^2-\overline{N}^2}{\omega k_z} w,
\Eeq
where the approximation is dropping terms of order $(k_zH)^{-1}$ and smaller.

The ratio of the $S'$ contribution to $\theta_T$ to the $\varpi'$ contribution to $\theta_T$ is
\Beq
\left|\frac{\overline{T}S'}{C_P}\right| \left|\frac{\varpi'}{\gamma}\right|^{-1} &=& \left(\frac{\overline{c_s}^2 k_{\perp}^2}{\omega^2}\right)\left(\frac{\partial_z\left(\frac{\overline{S}}{C_P}\right)}{k_z}\right). \label{entropy to pressure ratio}
\Eeq
Using  that $\overline{c_s}^2\sim \overline{N}^2 H^2$ and that $\omega\leq \overline{N}$ for gravity waves, one can show that this ratio is of order $k H$, which is assumed to be much larger than one\footnote{Near the cores of stars we instead have $\overline{c_s}^2\sim \overline{N}^2 r^2$.  In this case, the ratio is of order $k_z r\gg 1$.}.  Thus, $\theta_T$ is dominated by entropy diffusion.  Furthermore, the leading order (in $k_z H$) expressions for $\varpi'$ and $S'$ are out of phase, so upon volume integration, cross terms such as $\grad\varpi'\dot\grad S'$ are smaller than the leading order term $|\grad S'|^2$ by $(k_z H)^{-2}$.  Similarly, the volume average of terms like $S'\grad\overline{T}\dot\grad S'$ are also of order $(k_z H)^{-2}$ because $S'$ and $\partial_z S'$ are out of phase to leading order.

The leading order contributions to $\theta_T$ and $\theta_S$ are 
\Beq\label{theta T AN}
\theta_T & \approx & \frac{g\overline{\kappa_T}}{C_P^2\partial_z\overline{S}}\left|\grad S'\right|^2, \\
\theta_S & \approx & \frac{g\overline{\kappa_S}}{C_P\overline{T}\partial_z\overline{S}}\left|\grad S'\right|^2. \label{theta S AN}
\Eeq
For perturbations with $k_z H\gg 1$, we have that thermal diffusion and entropy diffusion are equivalent under the identification $\overline{\kappa_S} = \overline{\kappa_T}\overline{T}/C_P$.

A key assumption in the above argument is that $\omega\leq \overline{N}$, which was used to show that the ratio of the $S'$ contribution to $\theta_T$ to the $\varpi'$ contribution to $\theta_T$ (equation~\ref{entropy to pressure ratio}) is order $k H\gg 1$.  This is satisfied for gravity waves, but not for sound waves.  For sound waves, the first term on the RHS of equation~\ref{entropy to pressure ratio} is order one, so the $\varpi'$ contribution to $\theta_T$ is {\it larger} than the $S'$ contribution to $\theta_T$ by order $kH$.  This suggests that $T$- and $S$- diffusion give very different results for sound waves (see appendix~\ref{sec:sound waves}).

\subsection{FC \& PI equations}

Although we have only shown the results for the AN equations, similar results hold for the FC \& PI equations.  In appendix~\ref{sec:FC analytics}, we show that for the FC equations the $T$- and $S$- diffusion eigenvalue equations reduce to one another if we assume $k H\gg 1$ {\it and} that damping is weak.  Unlike the AN equations, there are differences between $T$- and $S$- diffusion for strongly damped modes (which are no longer waves).  We also derive analytic expressions for the damping rates.

In appendix~\ref{sec:PI analytics}, we carry out the same analysis as above for the PI equations.  It is straightforward to show that the PI equations will give the same damping rates using either $T$- or $S$- diffusion, assuming $k H \gg 1$.  This is because $T$- and $S$- diffusion in the PI equations only differ by a factor of $\overline{T}$, which can be absorbed into $\overline{\kappa}$ in the limit $k H \gg 1$.

\section{Steady Nonlinear Convective Solutions}\label{sec:steady convection}

We present nonlinear simulations of convection using the FC, PI, \& AN equations with both $T$- and $S$- diffusion models.  Near the onset of convection, unstable modes saturate as steady convective rolls.  For sufficiently small driving (${\rm Ra}$), these are stable.  However, as the driving of the system (${\rm Ra}$) increases, the rolls become unstable to oscillatory motions.  The accuracy of the stable convection states is an important nonlinear test of the different equation sets and thermal conduction models.  We restrict our investigation to 2D because the convection solutions are more susceptible to oscillatory motions in 3D.

We solve for convective steady states using Dedalus by integrating the equations of motion forward in time.  The background state is a polytrope (equations~\ref{polytrope T}-\ref{polytrope P}) with polytropic index
\Beq
n = 1.5 - \epsilon.
\Eeq
Recall that an adiabatic background has $n=1.5$; by setting $n$ slightly smaller than 1.5, we are imposing a slightly superadiabatic stratification.  We set $\epsilon=10^{-5}$.  This implies the background entropy gradient is
\Beq
\partial_z\left(\frac{\overline{S}}{C_P}\right) = - \frac{\epsilon}{L_z+1-z}.
\Eeq
Mixing length theory suggests that the Mach number squared is proportional to the entropy gradient, so we choose to fix the entropy gradient in order to fix the Mach number.  The convective steady states described in this paper have Mach numbers of about $10^{-3}$.

As above, we assume that the perturbation fields conduct heat differently from the background fields.  To satisfy thermal equilibrium, we assume that the background fields are acted upon by temperature diffusion with a constant $\overline{\kappa_T}$.  Again, it is consistent to use a different conduction model for the perturbations, provided that the perturbation fields never feed back onto the background fields.

For the perturbation fields, we use a constant diffusivity $\chi$,
\Beq
\vec{Q}_T' & = & - \chi_T \rho \grad T', \\
\vec{Q}_S' & = & - \chi_S \rho \overline{T} \grad \frac{S'}{C_P}.
\Eeq
Similarly, for viscosity, we use a constant diffusivity $\nu$,
\Beq
\mu = \rho \nu.
\Eeq
In both cases, $\rho$ is replaced by $\overline{\rho}$ for the AN equations.

We can now define the Rayleigh number: the ratio of driving to dissipation in the system,
\Beq\label{Ra}
{\rm Ra} = \frac{\Delta \overline{S} g L_z^3}{C_P \nu \chi},
\Eeq
where $\Delta \overline{S}/C_P = \epsilon \log(L_z+1)$ is the entropy jump across the domain (recall that in our non-dimensionalization $g=n+1$).  Below, we study how convective steady states vary as a function of ${\rm Ra}$.  All simulations have ${\rm Pr} = \nu/\chi = 1$.  We find that when ${\rm Pr}<1$, there are very few convective steady states, as the convection is strongly susceptible to oscillatory instabilities.  Thus, to compare the convection for ${\rm Pr}<1$ in different models would require a study of the (temporal) statistical properties of the flow, which is beyond the scope of this paper.  However, we do not expect our results to change qualitatively for low ${\rm Pr}$ convection (note that our linear wave results are at a ${\rm Pr}=0$).  Recent work by \citet{calkins14b} shows substantial differences between the onset of rapidly-rotating convection in the FC and AN equations with ${\rm Pr}<1$.  However, there is no indication that such differences persist in the non-rotating limit.

For simulations with ${\rm Pr}>1$, we do find some convective steady states.  In this case viscosity is a more dominant damping mechanism than thermal conduction, so we expect smaller differences between the conduction models.  For sufficiently high ${\rm Pr}$, the thermal boundary layers become unstable and there are no longer steady convective states.

To maintain the background entropy gradient, we use the boundary conditions $S'=0$ on the top and bottom.  Our other boundary conditions are $\vec{u}=0$ on the top and bottom, and all variables are periodic in the horizontal direction.  Our vertical boundary conditions are artificial---a more physical boundary condition would be to add stably stratified layers on either side of the convection zone.  We do not implement this type of background state because it greatly complicates the problem.  In real systems, convective plumes penetrate into the stably stratified regions, which in turn affects the whole convective state.  However, for stiff interfaces, the penetration is small \citep{RG05,brummell02}, suggesting that there might not be significant differences between a convective--radiative boundary and a solid wall.

Note that $w=0$ on the top and bottom are redundant equations for the horizontally averaged ($n=0$) mode when using the AN equations.  Furthermore, there is a gauge freedom in the definition of $\pi'$ for the PI equations and $\varpi'$ for the AN equations.  Thus, when using the PI or AN equations, for the horizontally averaged mode ($n=0$), we use the boundary conditions $w=0$ on the bottom boundary, and $\pi'$ or $\varpi'=0$ on the top boundary.  In fact, the PI equations are inconsistent for $\vec{u}=0$ on the top and bottom, and periodic in the horizontal directions.  This can be verified by integrating the constraint equation---the LHS is zero for these boundary conditions, but the RHS is generally non-zero.  Physically, this is because heating causes the fluid to expand, so there must be a way for the fluid to leave the box.

The simulations are run in a box with aspect ratio three, i.e., $L_x=3L_z$, with three different heights: $L_z=10, 30, 100$.  This corresponds to about $3, 5, 7$ density scale heights, respectively.  We use a resolution of 96 grid points in each direction, with 2/3 dealiasing (i.e., 64 modes).  We represent the solution as a Fourier series in the $x$ direction and a Chebyshev series in the $z$ direction.  For several cases, we ran with a resolution of 192 grid points (128 modes) in each direction, and found that the results were virtually identical.  To timestep the equations, we use an implicit--explicit, SBDF2 method \citep{ascher95,wr08}.  The time step is given by the smaller of the $0.3\vec{u}\dot\Delta \vec{x}^{-1}$, where $\Delta\vec{x}^{-1}$ is the inverse local grid spacing, and $5\times 10^{-5} \ \chi/L_z^2$.  The latter time scale almost always sets the time step, and was chosen to ensure the simulations are extremely well resolved temporally.

We do not base our CFL on the sound speed for the simulations of the FC equations (all equation sets are solved with the same time step size).  This is because we are able to implicitly timestep the sound waves, and are thus not limited by the sound speed CFL.  A similar approach was used by \citet{viallet11,viallet13}, although they use an iterative, nonlinear implicit solve, whereas we only treat linear terms implicitly.  This is significant as the typical rms Mach number of our nonlinear strongly stratified convection is $10^{-3}$, and explicitly following the sound waves would increase the computation cost by approximately $10^3$, which would make low-Mach number convection prohibitively expensive to simulate.  The greater complexity of the FC equations (see appendix~\ref{sec:dedalus equations}) makes them a factor of two slower than the AN simulations (the PI equations run at about the same speed as the FC equations).

The simulations are initialized by random, low amplitude density perturbations (FC \& PI equations) or entropy perturbations (AN equations).  The system is evolved forward in time until a convective steady state is found.  We assume we are in a steady state if the volume-averaged kinetic energy changes by less than a factor of $10^{-4}$ over one thousand iterations.  For higher ${\rm Ra}$, we sometimes did not find a steady state solution---instead, the system evolves into a periodically varying state.  In this case, we restart the simulation with different random initial conditions.  If several different random initial conditions lead to periodically varying states, we stop the search.  Of course, our limited search of initial conditions does not prove that there is no steady state for certain parameters, but it does suggest that the basin of attraction of a hypothetical steady state solution is likely limited.

\begin{figure*}
  \begin{center}
    \includegraphics[width=\linewidth]{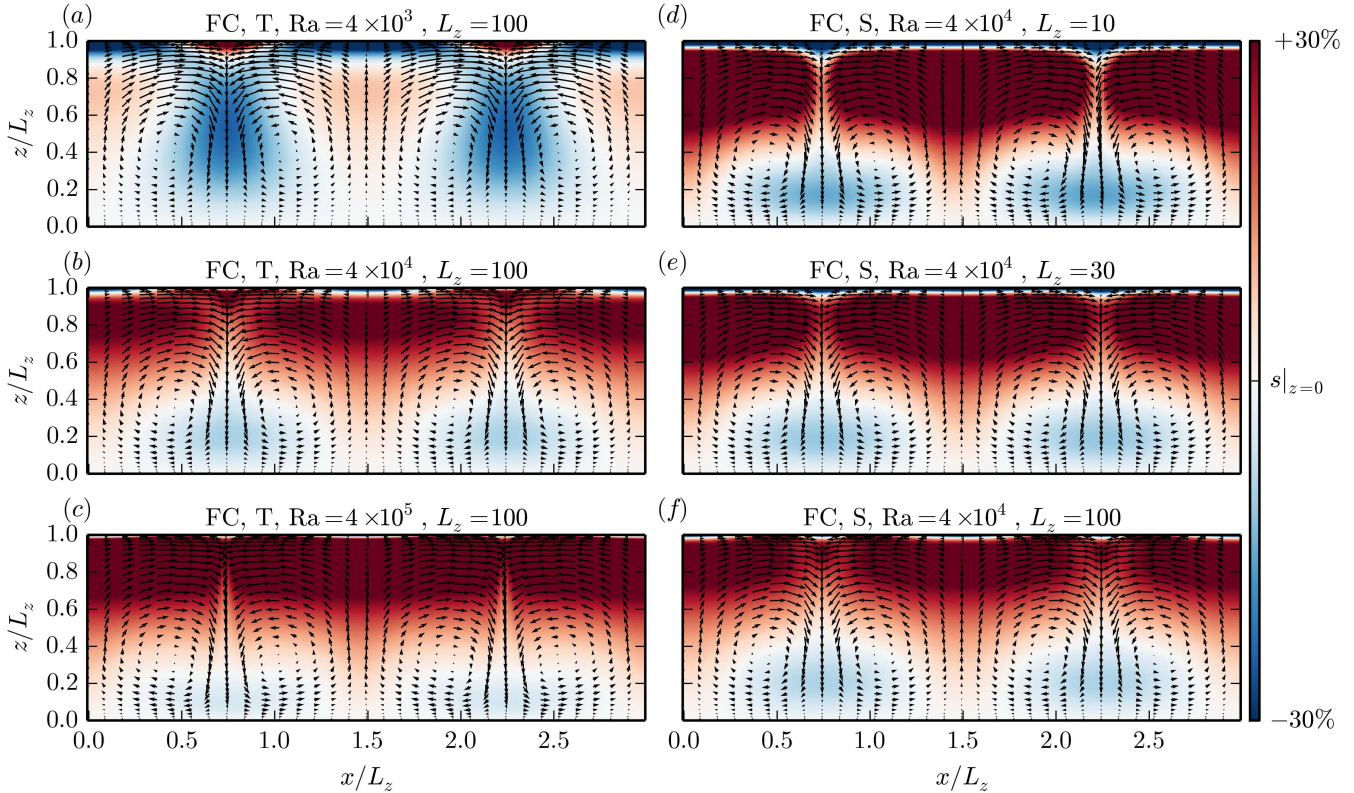}
  \end{center}
\caption{Steady convective states for the FC equations, varying the thermal conduction model, ${\rm Ra}$, and $L_z$.  The color depicts the total entropy (background plus perturbations)---the convective layers are fairly adiabatic, with a sharp boundary layer at the top of the domain.  The color scale is chosen such that white corresponds to the entropy at the bottom of the domain, and red (blue) is thirty percent greater (less) than this value.  The arrows show the flow field.  Each steady state is labelled with the equation set, the thermal conduction model ($T$- or $S$- diffusion), the Rayleigh number, and the vertical length of the domain.  In all cases, $L_x=3L_z$.  Panels ($a$)--($c$) use the FC equations with $T$-diffusion and $L_z=100$, but vary ${\rm Ra}$ from $4\times 10^3$ to $4\times 10^5$.  Panels ($d$)--($f$) use the FC equations with $S$-diffusion and ${\rm Ra}=4\times 10^4$, but vary $L_z$ from 10 to 100.
\label{fig:steady states parameter}
}
\end{figure*}

\begin{figure*}
  \begin{center}
    \includegraphics[width=\linewidth]{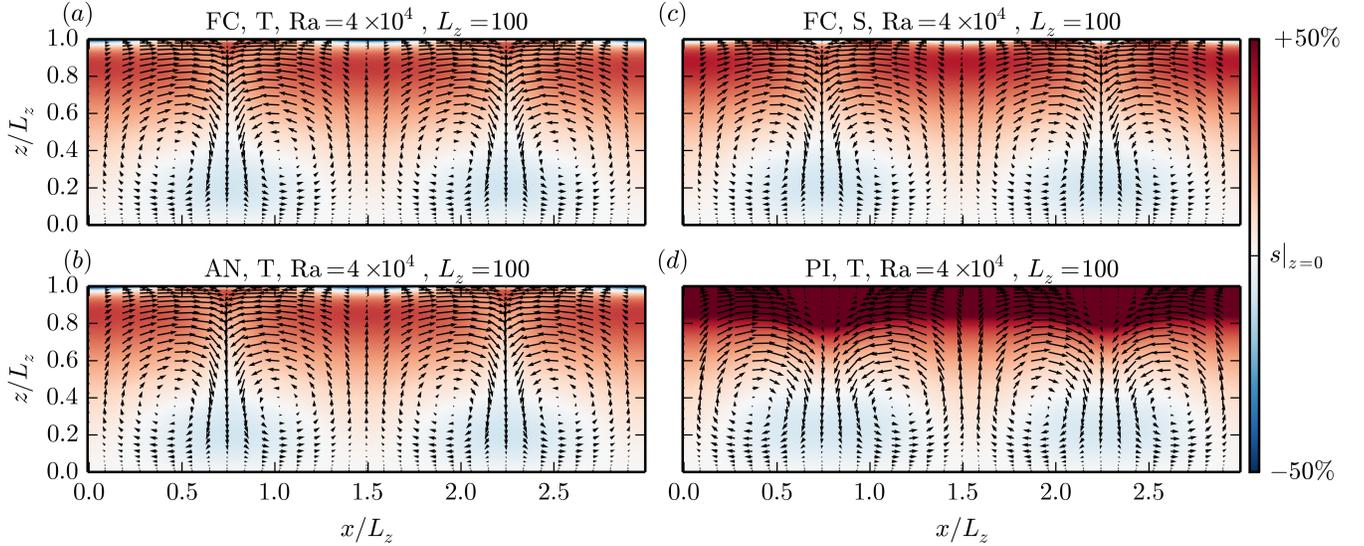}
  \end{center}
\caption{Steady convective states for different equations and thermal conduction models, with ${\rm Ra}=4\times 10^{4}$ and $L_z=100$.  The quantities plotted and labeling are the same as for figure~\ref{fig:steady states parameter}, except that the red (blue) colors correspond to fifty percent greater (less) than the entropy value at $z=0$.  In all cases, $L_x=3L_z$.  Panels ($a$) \& ($b$) use the $T$-diffusion with the FC \& AN equations, respectively.  These plots are virtually identical.  Panel ($c$) shows the convective steady state for the FC equations with $S$-diffusion; the results for the AN \& PI equation equations with $S$-diffusion are virtually identical and not shown. Panel ($d$) shows the convective steady state for the PI equations with $T$-diffusion.
\label{fig:steady states equation}
}
\end{figure*}

Sometimes we found multiple convective steady states.  The most common state consists of two pairs of convective rolls.  To facilitate comparison between different states, we only consider steady states consisting of two pairs of convective rolls (see figures~\ref{fig:steady states parameter} \& \ref{fig:steady states equation}).  If we changed the horizontal periodicity of the domain, the aspect ratio of the convective cells would likely stay close to the aspect ratio described here, as this is the preferred aspect ratio of the system.

Figure~\ref{fig:steady states parameter} plots several convective steady states using the FC equations.  We vary ${\rm Ra}$ in panels ($a$)--($c$), fixing $L_z=100$ and using $T$-diffusion.  As ${\rm Ra}$ increases, the boundary layer at the top of the domain decreases in size, and the flow and entropy become more asymmetric.  The down flows between convective rolls become sharper as ${\rm Ra}$ increases.  At ${\rm Ra}$ somewhat higher than $4\times 10^5$, the down flows become so sharp that they become unstable to an oscillatory instability.

In the bulk of the fluid, up flows carry high entropy fluid, and down flows carry low entropy fluid.  However, things become more complicated in the upper boundary layer for highly stratified convection, where sometimes high entropy fluid rests above down flows, and low entropy fluid rests above up flows (see panel ($a$)).  This could be due to two effects.  First, viscous heating increases the entropy near the down flows where there are sharp velocity gradients.  Second, due to large stratification near the top of the box, up flows produce diverging flows which dilute entropy, whereas down flows produce converging flows which concentrate entropy.

Panels ($d$)--($f$) show convective steady states with $S$-diffusion for fixed ${\rm Ra}=4\times 10^4$, but with increasing $L_z$.  Increasing $L_z$ also decreases the thickness of the boundary layer (relative to the box size) at the top of the domain.  Note that as $L_z$ increases, the critical ${\rm Ra}$ (at which convection begins) also increases.  Thus, ${\rm Ra}/{\rm Ra}_c$ is decreasing with increasing $L_z$, which might naively lead one to believe that the boundary layer thickness should {\it increase} with increasing $L_z$, the opposite of what we find.

However, these convective steady states are in some ways consistent with the idea that convection only occurs on the local scale height.  In all cases, the local scale height is $\approx 1$ near the top of the box.  This corresponds to $1/10$ of the domain when $L_z=10$, but $1/100$ of the domain when $L_z=100$.  Thus, the upper boundary layer might be influenced by the local scale height near the top of the box.  The convective rolls also seem to have size $\sim L_z$, which is similar to the scale height at the bottom of the domain.  However, there are no features on intermediate scales, even for $L_z=100$ which contains seven density scale heights.

In figure~\ref{fig:steady states equation}, we vary the equations and thermal conduction model, fixing ${\rm Ra}=4\times 10^{4}$ and $L_z=100$.  Panels ($a$) \& ($c$) correspond to panels ($b$) \& ($f$) in figure~\ref{fig:steady states parameter}.  The FC \& AN equations with $T$-diffusion (panels ($a$) \& ($b$)) look virtually identical.  There are slight differences between the FC equations with $T$- and $S$- diffusion (panels ($a$) \& ($c$))---the upper boundary layer is slightly smaller for $S$-diffusion.

However, there are substantial differences between the steady state for the PI equations with $T$-diffusion (panel ($d$)) and the other three steady states.  The entropy variation with height is very different for the PI steady state---the entropy is much larger at the top of the box than the rest of the domain.  It might appear that the convective steady state does not satisfy the $S'=0$ boundary condition at the top of the domain.  However, this is only because there is an extremely thin boundary layer at the top of the domain which is well resolved in the simulation, but is smaller than the resolution of the image.  Also, the flow pattern looks very different, with less asymmetry between the up flows and down flows than in the FC \& AN simulations.  The convective steady states were virtually identical for the FC \& AN equations using $T$-diffusion, and all three equation sets with $S$-diffusion.  The only equations which show strong differences from the others are the PI equations with $T$-diffusion.

\begin{figure}
  \begin{center}
    \includegraphics[width=9cm]{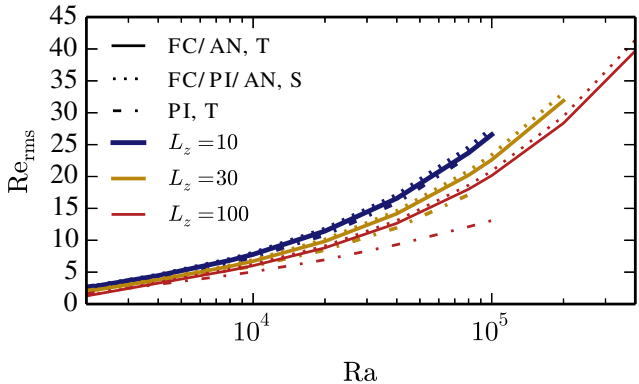}
  \end{center}
\caption{${\rm Re}_{\rm rms}$ (equation~\ref{rms Re}) of convective steady states as a function of ${\rm Ra}$ (equation~\ref{Ra}) for different equation sets, thermal conduction models, and box sizes $L_z$.  Solid lines show results for the FC \& AN equations with $T$-diffusion, dotted lines show results for the FC, PI, \& AN equations with $S$-diffusion, and dot-dashed lines show results for the PI equations with $T$-diffusion.  Blue, yellow, and red lines show results for $L_z=10$, $30$, and $100$ respectively (recall that $L_x=3L_z$).  The different equation sets which have been grouped together have differences in ${\rm Re}_{\rm rms}$ of less than 1\% (and typically less than 0.01\%) for each ${\rm Ra}$ and $L_z$ shown.  Although the $T$- and $S$- diffusion models track each other fairly well for the FC \& AN equations, the PI equations with $T$-diffusion have convective steady states with low ${\rm Re}_{\rm rms}$, especially for highly stratified domains.
\label{fig:Re}
}
\end{figure}

To more quantitatively compare the different convective steady states, we plot the rms ${\rm Re}$ in figure~\ref{fig:Re}.  We define the rms ${\rm Re}$ to be
\Beq\label{rms Re}
{\rm Re}_{\rm rms} = \frac{\sqrt{\langle |\vec{u}|^2\rangle} L_z}{\nu}.
\Eeq
Note that we were not able to find convective steady states for high ${\rm Ra}$ for some of the low $L_z$ boxes, and with the PI equations with $T$-diffusion.

The $T$- and $S$- diffusion models produced very similar results for the FC \& AN equations.  For all three $L_z$, the difference in ${\rm Re}_{\rm rms}$ at the lowest ${\rm Ra}$ was less than 1\%, and at the highest ${\rm Ra}$ was $\approx 4\%$.  In contrast, the PI equations with $T$-diffusion has convective steady states which are rather different (note that the PI equations with $S$-diffusion give results practically indistinguishable from the FC or AN equations with $S$-diffusion).  Although the differences in ${\rm Re}_{\rm rms}$ between the FC and PI equations with $T$-diffusion are $< 5\%$ for all ${\rm Ra}$ for $L_z=10$, at high ${\rm Ra}$, the differences grow to $15\%$ for $L_z=30$ and $35\%$ for $L_z=100$.  These results are consistent with the similarity and differences between the convective steady states shown in figure~\ref{fig:steady states equation}.

It might seem odd that the PI equations seem to be substantially different from the FC \& AN equations with $T$-diffusion, but virtually identical to the FC \& AN equations with $S$-diffusion.  The difference lies in the PI equation of state (equation~\ref{PI equation of state}), in which $P$ is replaced by $\overline{P}$.  For low Mach number convection, $\rho'/\overline{\rho}$, $P'/\overline{P}$, and $S'/C_P$ are all $\mathcal{O}({\rm Ma}^2)$.  However, the PI equation of state assumes that $\rho'/\overline{\rho}$ and $S'/C_P$ are $\mathcal{O}(1)$, but that $P'/\overline{P}\sim\mathcal{O}({\rm Ma}^2)$.  Thus, the PI equation of state introduces inaccuracies in thermodynamic variables.

When using $S$-diffusion, the equation of state is in some sense ``not used.''  Summing the PI continuity equation (\ref{PI full continuity}) and the PI constraint equation (\ref{PI full constraint}), and using the PI equation of state (\ref{PI equation of state}), one can show that $S'_{\rm PI}$ and $S'_{\rm AN}$ satisfy the same equation.  However, when using $T$-diffusion, the equation of state must be used to relate $T'_{\rm PI}$ to other thermodynamic quantities.  In this case, there are differences in $T'_{\rm PI}$ and $T'_{\rm AN}$ because the latter depends on the pressure perturbation.

\section{Conclusion}\label{sec:conclusion}

This paper examines the differences between the temperature diffusion and entropy diffusion models of thermal conduction, for three different equations sets: the fully compressible equations, the pseudo-incompressible equations, and the anelastic equations.  We study both damping rates of linear internal gravity wave modes, and the properties of low Rayleigh number convective steady states.

Overall, we find little difference between temperature diffusion and entropy diffusion, provided that the conductivities are related by
\Beq
\overline{\kappa_T} = \overline{T}\overline{\kappa_S}/C_P.
\Eeq
Using a different relation between $\overline{\kappa_T}$ and $\overline{\kappa_S}$ will cause differences between the temperature and entropy diffusion models, just as there are differences between temperature diffusion using two different conductivities.  Although entropy diffusion could lead to non-monotonicity of the total entropy, in practice, it generally does not, as the $\grad S$ field is often aligned with the $\grad T$ field (section~\ref{sec:damping rates}).  The only way to ensure that entropy will increase monotonically is to use temperature diffusion \citep{ll59}.

Temperature and entropy diffusion give the same linear internal gravity wave damping rates for all three equation sets, provided that $k H \gg 1$, i.e., that the wavelength is shorter than the density scale height.  For $k H \lesssim 1$, we find modest differences between damping rates for different equation sets and thermal conduction models (see figure~\ref{fig:eigenvalues}).  The longest wavelength modes we studied ($\approx 6.6$ density scale heights) have damping rate errors of $\sim 20\%$ for the anelastic equations (with either thermal conduction model) and pseudo-incompressible equations with entropy diffusion, but errors of $\sim 50\%$ for the pseudo-incompressible equations with temperature diffusion.  We believe the large errors in the pseudo-incompressible equations with temperature diffusion are due to inaccuracies in the pseudo-incompressible equation of state (which assumes that the pressure perturbations are much smaller than the density and entropy perturbations, and can be dropped).

We also calculate convective steady states using Dedalus (section~\ref{sec:steady convection}).  The flexibility of Dedalus allows us to study the fully compressible, anelastic, and pseudo-incompressible equations all within the same framework.  Furthermore, by implicitly timestepping sound waves, we are able to take the same time step in fully compressible calculations as the anelastic and pseudo-incompressible calculations (despite having Mach numbers of $10^{-3}$).  Because the implementation of the fully compressible equations is more complicated than the implementation of the anelastic equations, we find that the fully compressible simulations run about half as fast as the anelastic simulations.  The pseudo-incompressible simulations run at about the same speed as the fully compressible simulations.

For Rayleigh numbers above the instability threshold of convection, but below the onset of oscillatory instabilities, we find many convective steady states for box sizes ranging from three density scale heights to seven density scale heights.  The convective steady states are essentially identical for the fully compressible equations and anelastic equations with temperature diffusion; similarly, the convective steady states are essentially identical for the all three equation sets using entropy diffusion.  Furthermore, these two classes of steady states are very similar (figure~\ref{fig:Re}).  However, the pseudo-incompressible equations with temperature diffusion find convective steady states with much lower rms Reynolds number than the other equations, and with very different convection patterns (figure~\ref{fig:steady states equation}).  We again attribute the difference to the incorrect equation of state, which does not correctly calculate the temperature perturbation for low Mach number convection.  The differences are largest for the most strongly stratified domains, as the convective rolls have longest wavelengths ($kH \lesssim 1$).

In a similar analysis, \citet{calkins14} calculate the critical Rayleigh number for the onset of convection for the fully compressible equations (with temperature diffusion) and the anelastic equations with both temperature and entropy diffusion, also including the effects of rotation.  They also find that the anelastic and fully compressible equations with temperature diffusion give nearly identical results in the low Mach number limit (when the background is very close to adiabatic).  They find much larger differences between entropy and temperature diffusion than we do.  This is likely because their entropy diffusion model diffuses $\overline{T} S'$, unlike our own which diffuses $S'$.  Diffusing $\overline{T} S'$ is equivalent to using temperature diffusion in the pseudo-incompressible equations, which we have shown can produce substantial errors for highly stratified domains.

Although we find only minor differences between temperature and entropy diffusion for linear internal waves and low Rayleigh number convective steady states, there is no guarantee that these different thermal conduction models will continue to give similar results in the strongly nonlinear regime.  In the future, we plan to investigate the effects of different thermal conduction models on strongly nonlinear wave breaking and high Rayleigh number convection.  Our present results show that the differences between temperature and entropy diffusion in convective steady states grows as the Rayleigh number increases, with entropy diffusion overestimating the velocities in the convective steady states.  Perhaps this indicates that at the very high Rayleigh numbers of stellar convection, there are substantial and important differences between temperature and entropy diffusion.

\acknowledgements{We thank Mike Calkins, Nick Featherstone, Kyle Augustson, Eliot Quataert, and Keith Julian for helpful discussions on different conduction models.  DL is supported by a Hertz Foundation Fellowship, the National Science Foundation Graduate Research Fellowship under Grant No. DGE 1106400, a Kavli Institute for Theoretical Physics Graduate Student Fellowship, and partially by the Schneider Chair in Physics to Eliot Quataert. BPB acknowledges support from NSF Astronomy and Astrophysics postdoctoral fellowship AST 09-02004 and a KITP postdoctoral fellowship (Grant No. NSF PHY11-25915). EGZ acknowledges support from NSF via the Center for Magnetic Self-Organization in Laboratory and Astrophysical Plasmas (CMSO) (PHY08-21899), from NASA under TCAN grant number NNX14AB53G, and the hospitality of KITP. KJB is supported by a MIT Kavli Institute Graduate Fellowship and a National Science Foundation Graduate Research Fellowship under Grant No. 1122374.  JSO acknowledges support from NSF Grant AST10-09802.  GMV is generously supported by the Australia Research Council, project number DE140101960. Part of this work was completed at the Kavli Institute of Theoretical Physics programs on Wave-Mean Flow Interaction, and Star Formation (Grant No. NSF PHY11-25915).}

\appendix

\section{Eigenvalue Equations and Damping Rates for Fully Compressible Equations}\label{sec:FC analytics}

The eigenvalue equations for the FC equations are somewhat more complicated than for the AN equations.  We again drop all terms of order $(k H)^{-1}\ll 1$.  The eigenvalue equation for $T$-diffusion is
\Beq
&&\left[\omega^2\left(1 - i\frac{\overline{\kappa_T}\nabla^2}{\omega C_P}\right)^2 \nabla^2 + \frac{\omega^4}{\overline{c_s}^2} \left(1 - i\frac{\chi_T\nabla^2}{\overline{\rho}C_P\omega}\right)\left(1-i\frac{\gamma\overline{\kappa_T}\nabla^2}{\overline{\rho} C_P \omega}\right) \right. \nonumber \\
&&\quad \left. - \frac{g}{C_P} \left(1- i\frac{\overline{\kappa_T}\nabla^2}{\overline{\rho} C_P \omega}\right)\left(\partial_z\overline{S}\left(\nabla_{\perp}^2 + \frac{\omega^2}{\overline{c_s}^2}\right) +i\omega \frac{\gamma-1}{\overline{c_s}^2}\frac{\overline{\kappa_T}}{\overline{\rho}}\nabla^2 \partial_z\right) \right] w = 0.
\Eeq
The eigenvalue equation for $S$-diffusion is
\Beq
\left[\omega^2 \left(1 - i\frac{\overline{\kappa_S}\nabla^2}{\overline{\rho}\overline{T} \omega}\right)\left(\nabla^2  + \frac{\omega^2}{\overline{c_s}^2}\right) - g \frac{\partial_z\overline{S}}{C_P} \left(\nabla_{\perp}^2+ \frac{\omega^2}{\overline{c_s}^2}\right)\right] w = 0.
\Eeq
Unlike the eigenvalue equations for the AN equations, these are not equivalent.  This is partially because sound waves are affected differently by temperature and entropy diffusion.  IGWs with $k H \gg 1$ have the property
\Beq
k^2 \gg \frac{\omega^2}{\overline{c_s}^2}.
\Eeq
If we use this relation, the eigenvalue equation with $T$-diffusion is
\Beq
\left[\omega^2 \left(1 - i \frac{\overline{\kappa_T}\nabla^2}{\overline{\rho}C_P \omega}\right)\nabla^2 - \frac{g}{C_P} \left(\partial_z\overline{S}\nabla_{\perp}^2 + i\omega \frac{\gamma-1}{\overline{c_s}^2}\frac{\overline{\kappa_T}}{\overline{\rho}} \nabla^2\partial_z\right)\right] w = 0,\label{FC T IGW}
\Eeq
while the eigenvalue equation with $S$-diffusion is
\Beq
\left[\omega^2 \left(1 - i\frac{\overline{\kappa_S}\nabla^2}{\overline{\rho}\overline{T} \omega}\right)\nabla^2 - g \frac{\partial_z\overline{S}}{C_P} \nabla_{\perp}^2\right] w = 0.
\Eeq
These equations are still not equivalent, due to the last term in equation~\ref{FC T IGW}.  It cannot be shown to be small in comparison to $\partial_z\overline{S}\nabla_{\perp}^2$ unless $\overline{\kappa_T}$ is assumed to be small.  For weakly damped waves, the eigenvalue equations are equivalent under the identification $\overline{\kappa_S}=\overline{T}\overline{\kappa_T}/C_P$.  However, for strongly damped modes (which are no longer wave-like), we expect larger differences.

One can show that the perturbation energy for the FC equations is
\Beq
E = \frac{1}{2}\overline{\rho}|\vec{u}|^2 + \frac{1}{2}\frac{g\overline{\rho}}{C_P\partial_z\overline{S}} S'^2 + \frac{1}{2}\frac{\overline{\rho}}{\overline{c_s}^2} \varpi'^2,
\Eeq
and the $T$- and $S$- diffusion damping terms are
\Beq
\theta_T & = & \overline{\kappa_T} \grad\left(\frac{g S'}{C_P\overline{T} \partial_z\overline{S}} + \frac{\varpi'}{\overline{T}C_P}\right)  \dot\grad \left(\overline{T} \frac{S'}{C_P} + \frac{\gamma-1}{\gamma}\varpi' \right), \\
\theta_S & = & \overline{\kappa_S}\grad\left(\frac{g S'}{C_P\overline{T} \partial_z\overline{S}} + \frac{\varpi'}{\overline{T}C_P}\right)\dot\grad S'.
\Eeq
For weak dissipation, both $S'$ and $\varpi'$ can be related to $w$ by
\Beq
S' & = & -i\frac{w\partial_z\overline{S}}{\omega}, \\
\varpi' & = & i\omega \frac{\partial_z w + \frac{1}{\gamma}\left(\partial_z\log\overline{p}\right)w}{\frac{\omega^2}{\overline{c_s}^2}+k_{\perp}^2} \approx \frac{\omega^2-\overline{N}^2}{\omega k_z} w,
\Eeq
where the approximation is dropping terms of order $(k_zH)^{-1}$ and smaller.  This implies the $S'$ contribution is much larger than the $\varpi'$ contribution to $\theta_T$ and $\theta_S$,
\Beq
\left|\frac{\overline{T}S'}{C_P}\right| \left|\frac{\varpi'}{\gamma}\right|^{-1} &=& \left(\frac{\overline{c_s}^2 k_z^2}{\omega^2-\overline{N}^2}\right)\left(\frac{\partial_z\left(\frac{\overline{S}}{C_P}\right)}{k_z}\right), \\
\left|\frac{g S'}{\partial_z\overline{S}}\right| \left|\varpi'\right|^{-1}& =& \frac{gH k_z^2}{\omega^2-\overline{N}^2} \frac{1}{k_z H}.
\Eeq
Using the IGW eigenvalue equation, and that $\overline{c_s}^2\sim \overline{N}^2 H^2$, one can show that both these terms are order $(k H)^{-1}$.

The leading order contributions to $\theta_T$ and $\theta_S$ are 
\Beq\label{theta T FC}
\theta_T & \approx & \frac{g\overline{\kappa_T}}{C_P^2\partial_z\overline{S}}\left|\grad S'\right|^2, \\
\theta_S & \approx & \frac{g\overline{\kappa_S}}{C_P\overline{T}\partial_z\overline{S}}\left|\grad S'\right|^2. \label{theta S FC}
\Eeq
Thus, for perturbations with $k_z H\gg 1$ and weak dissipation, we have that thermal diffusion and entropy diffusion are equivalent under the identification $\overline{\kappa_S} = \overline{\kappa_T}\overline{T}/C_P$.

\section{Eigenvalue Equations and Damping Rates for Pseudo-Incompressible Equations}\label{sec:PI analytics}

Recall that for the linearized PI equations $T'=\overline{T}S'/C_P$.  This makes temperature and entropy conduction extremely similar.  In the limit $kH\gg 1$, the eigenvalue equation for $T$-diffusion is
\Beq
\left[\omega^2\left(1 - i\frac{\overline{\kappa_T}}{\overline{\rho}C_P \omega}\nabla^2\right)\nabla^2 - g \partial_z\frac{\overline{S}}{C_P}\nabla_{\perp}^2 \right]w = 0.
\Eeq
The eigenvalue equation for $S$-diffusion is
\Beq
\left[\omega^2\left(1 - i\frac{\overline{\kappa_S}}{\overline{\rho}\overline{T} \omega}\nabla^2\right)\nabla^2 - g \partial_z\frac{\overline{S}}{C_P}\nabla_{\perp}^2 \right]w = 0.
\Eeq
These are equal to each other when $\overline{\kappa_S}=\overline{\kappa_{T}}\overline{T}/C_P$.  This shows that $T$- and $S$- diffusion are equivalent for the PI equations as long as $k H \gg 1$, irrespective of the strength of thermal conduction.

The expression for the perturbation energy is
\Beq
E = \frac{1}{2}\overline{\rho}\left|\vec{u}\right|^2 + \frac{1}{2}\frac{g\overline{\rho}}{C_P\partial_z\overline{S}} S'^2,
\Eeq
and the thermal and entropy damping terms are
\Beq
\theta_T & = & \overline{\kappa_T} \grad\left(\frac{g S'}{C_P\overline{T} \partial_z\overline{S}} + \frac{\varpi'}{\overline{T}C_P}\right) \dot\grad \left(\overline{T} \frac{S'}{C_P}  \right), \\
\theta_S & = & \overline{\kappa_S}\grad\left(\frac{g S'}{C_P\overline{T} \partial_z\overline{S}} + \frac{\varpi'}{\overline{T}C_P}\right)\dot\grad S'.
\Eeq
This is very similar to the FC expressions, except that $T'=\overline{T} S'/C_P$, since there is no $\varpi'$ term in the equation of state.  In the limit of $k H\gg 1$, these two expressions are equivalent, since
\Beq
\grad\left(\overline{T} \frac{S'}{C_P}  \right)\approx  \frac{\overline{T}}{C_P} \grad  \left(S'  \right).
\Eeq

\section{Thermal Conduction and Sound Waves}\label{sec:sound waves}

In this paper, we focus on the differences in IGW damping rates between equation sets and thermal conduction models.  However, the FC equations also admit sound waves.  Here we demonstrate, using Dedalus, the substantial differences in sound wave damping when using either $T$- or $S$- diffusion.

We solve the linear FC equations for the same background and parameters as in section~\ref{sec:IGW numerics}.  We check the damping rate of each mode against the analytic result given in appendix~\ref{sec:FC analytics}.  The sound waves are much harder to resolve than IGWs---thus, some of our sound wave  damping rates disagree with the analytic damping rates by as much as 20\% (with a vertical resolution of 256 modes).  However, the differences between $T$- and $S$- diffusion are much larger than this, so we have not repeated the calculation at higher resolution to reduce the errors.

\begin{figure*}
  \begin{center}
    \includegraphics[width=\linewidth]{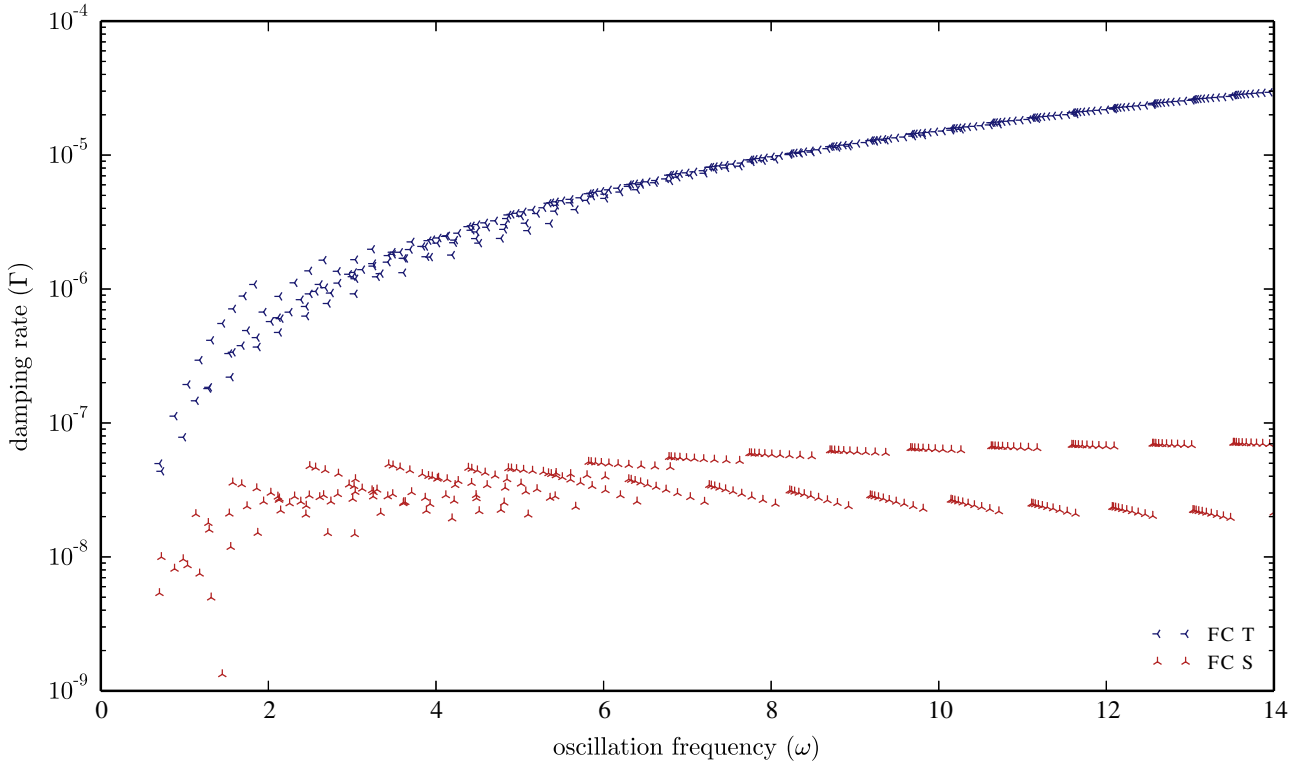}
  \end{center}
\caption{Damping rates and oscillation frequencies of sound wave modes of the FC equations, using either $T$- or $S$- diffusion.  The first ten radial and horizontal modes are shown.  Each cluster of modes (most visible at high oscillation frequency) corresponds to a single horizontal mode number.  The oscillation frequency increases with increasing wavenumber.
\label{fig:sound}}
\end{figure*}

In figure~\ref{fig:sound}, we plot the damping rates and oscillation frequencies for sound wave modes, using either $T$- or $S$- diffusion.  Although the damping rate increases with increasing $k$ (and increasing oscillation frequency) for $T$-diffusion, the damping rate stays about constant for $S$-diffusion.  The oscillation frequencies between the modes agree well because they are only weakly damped.

For sound waves, the damping rate using $S$-diffusion becomes increasing inaccurate as $k$ increases.  This is the opposite result as for IGWs.  Although the dominant contribution to $T'$ is $S'$ for IGWs, the dominant contribution to $T'$ is the pressure perturbation $\pomega'$ for sound waves.  In section~\ref{sec:IGW analytics}, we show that the ratio of the $\pomega'$ contribution to $T'$, to the $S'$ contribution to $T'$, decreases with increasing $k$ if the modes follow the IGW dispersion relation, but increases with increasing $k$ if the modes follow the sound wave dispersion relation.

Physically, this is because sound waves are almost adiabatic waves.  In the absence of gravity, they are completely adiabatic.  On the other hand, as they are pressure-driven waves, they have large pressure perturbations, which correspond to large temperature perturbations.  Thus, sound waves are much more efficiently damped with $T$-diffusion than $S$-diffusion.

\section{Equation Implementation in Dedalus}\label{sec:dedalus equations}

Dedalus solves systems of equation which are first order in $z$.  Time derivatives are here discretized using an implicit--explicit scheme; equations are written such that terms on the LHS of the equals sign are temporally discretized implicitly (i.e., the term is evaluated in the future), and terms on the RHS of the equals sign are temporally discretized explicitly (i.e., the term is only evaluated in the present and/or past).  Only linear terms can be treated implicitly, although they are not required to be.

\subsection{Fully Compressible Equations}

We implement the fully compressible equations with temperature diffusion as
\Beq
&&\partial_t w + \partial_z T' + \overline{T} \partial_z\Upsilon' + T'\partial_z\log\overline{\rho} - \nu\left[\partial_x^2 w + \partial_z w_z + 2\partial_z \log\overline{\rho} w_z + \frac{1}{3} \left(\partial_x u_z + \partial_z w_z\right) - \frac{2}{3}\partial_z\log\overline{\rho}\left(\partial_x u + w_z\right)\right] \nonumber \\
&& \quad \quad \quad = - T' \partial_z\Upsilon' - u\partial_x w - w w_z +\nu \left[ u_z\partial_x\Upsilon' + 2 w_z \partial_z\Upsilon' +\partial_x w \partial_x\Upsilon' - \frac{2}{3}\partial_z\Upsilon'\left(\partial_x u + w_z\right)\right], \label{FC vertical dedalus} \\
&&\partial_t u + \partial_xT' + \overline{T}\partial_x\Upsilon' - \nu\left[ \partial_x^2 u +\partial_z u_z + \partial_z\log\overline{\rho} \left(u_z + \partial_x w \right) + \frac{1}{3} \left(\partial_x^2 u + \partial_x w_z\right)\right] \nonumber \\
&&\quad \quad \quad = - T'\partial_x\Upsilon' - u\partial_x u - w u_z +\nu\left[2\partial_x u \partial_x\Upsilon' + \partial_x w \partial_z\Upsilon' + u_z \partial_z\Upsilon' - \frac{2}{3} \partial_x\Upsilon'\left(\partial_x u + w_z\right)\right], \label{FC horizontal dedalus} \\
&&\partial_t\Upsilon' + w\partial_z\log\overline{\rho} + \partial_x u + w_z = -u\partial_x\Upsilon' - w \partial_z\Upsilon' \label{FC continuity dedalus} \\
&&\partial_t T' + w\partial_z \overline{T} + \left(\gamma - 1\right) \overline{T}\left(\partial_x u + w_z\right) - \frac{\chi}{C_V}\left(\partial_x^2 T' - \partial_z\tilde{Q}'_z - \tilde{Q}'_z\partial_z\log\overline{\rho}\right) \nonumber \\
&&\quad \quad \quad = - u\partial_x T' - w\partial_z T' - \left(\gamma - 1\right) T'\left(\partial_x u + w_z\right) + \frac{\chi}{C_V}\left(\partial_x T'\partial_x\Upsilon' - \tilde{Q}'_z\partial_z\Upsilon'\right) \nonumber \\
&&\quad \quad \quad \quad + \frac{\nu}{C_V}\left[2\left(\partial_x u\right)^2 + \left(\partial_x w\right)^2 + u_z^2 + 2 w_z^2 + 2 u_z\partial_x w - \frac{2}{3}\left(\partial_x u + w_z\right)^2\right], \label{FC temperature dedalus} \\
&&\tilde{Q}'_z + \partial_z T' = 0, \label{FC heat flux dedalus}\\
&& \frac{S'}{C_P} - \frac{T'}{\gamma \overline{T}} + \frac{1}{C_P} \Upsilon' = \frac{1}{\gamma}\left[\log\left(1 + \frac{T'}{\overline{T}}\right) - \frac{T'}{\overline{T}}\right], \label{FC EoS dedalus} \\
&&w_z - \partial_z w = 0, \label{FC wz dedalus} \\
&&u_z - \partial_z u = 0. \label{FC uz dedalus}
\Eeq
In these equations, $u$ and $w$ are the horizontal and vertical velocity, respectively.  The temperature field is decomposed as $T=\overline{T} + T'$, the density field is decomposed as $\log\rho = \log\overline{\rho} + \Upsilon'$, and the entropy field is decomposed as $S = \overline{S} + S'$.  The normalized vertical heat flux $\tilde{Q}'_z$ is the vertical component of the heat flux divided by $\chi \rho$.  Equations~\ref{FC vertical dedalus} \& \ref{FC horizontal dedalus} are the vertical and horizontal momentum equation respectively, equation~\ref{FC continuity dedalus} is the continuity equation, while equation~\ref{FC temperature dedalus} is the equation for temperature.  Equation~\ref{FC heat flux dedalus} defines the heat flux, equation~\ref{FC EoS dedalus} is the fully nonlinear equation of state, while equations~\ref{FC wz dedalus} \& \ref{FC uz dedalus} define quantities which have second vertical derivatives within our first order system.  These are equivalent to equations~\ref{FC momentum nonlinear}--\ref{FC entropy nonlinear}.  Note that the entropy perturbation solved for in the equation of state (\ref{FC EoS dedalus}) is only used to enforce the $S'=0$ boundary condition.  The momentum equations and the temperature equation have nonlinear diffusion terms which are treated explicitly.

For the FC equations with entropy diffusion, equations~\ref{FC temperature dedalus} \& \ref{FC heat flux dedalus} are replaced by
\Beq
&&\partial_t T' + w\partial_z \overline{T} + \left(\gamma - 1\right) \overline{T}\left(\partial_x u + w_z\right) - \frac{\chi}{C_V}\left(\overline{T}\partial_x^2\frac{S'}{C_P} - \partial_z\tilde{Q}'_z - \tilde{Q}'_z\partial_z\log\overline{\rho}\right) \nonumber \\
&&\quad \quad \quad = - u\partial_x T' - w\partial_z T' - \left(\gamma - 1\right) T'\left(\partial_x u + w_z\right) + \frac{\chi}{C_V}\left(\overline{T}\partial_x \frac{S'}{C_P}\partial_x\Upsilon' - \tilde{Q}'_z\partial_z\Upsilon'\right) \nonumber \\
&&\quad \quad \quad \quad + \frac{\nu}{C_V}\left[2\left(\partial_x u\right)^2 + \left(\partial_x w\right)^2 + u_z^2 + 2 w_z^2 + 2 u_z\partial_x w - \frac{2}{3}\left(\partial_x u + w_z\right)^2\right], \\
&& \tilde{Q}'_z + \overline{T}\partial_z\frac{S'}{C_P} = 0.
\Eeq

Note that $\chi$, $\nu$, and $\gamma$ are assumed to be constant.  In both conduction models, we solve for the variables $u, u_z, w, w_z, \Upsilon', T', S', \tilde{Q}'_z$.

\subsection{Pseudo-Incompressible Equations}

Our implementation of the PI equations with temperature diffusion is
\Beq
&&\partial_t w + \partial_z \pomega' + \overline{\pomega}\partial_z\Upsilon' - \pomega'\partial_z\frac{\overline{S}}{C_P} - \nu\left[\partial_x^2 w + \partial_z w_z + 2\partial_z\log\overline{\rho} w_z + \frac{1}{3}\left(\partial_x u_z + \partial_z w_z\right) - \frac{2}{3}\partial_z\log\overline{\rho}\left(\partial_x u +w_z\right) \right] \nonumber \\ 
&& \quad \quad \quad= -\pomega'\partial_z\Upsilon' - u\partial_x w - w w_z+ \nu\left[u_z \partial_x\Upsilon' + 2 w_z \partial_z\Upsilon' + \partial_x w \partial_x\Upsilon' - \frac{2}{3}\partial_z\Upsilon'\left(\partial_x u + w_z\right)\right], \label{PI vertical dedalus} \\
&&\partial_t u + \partial_x\pomega' + \overline{\pomega}\partial_x\Upsilon' - \nu\left[\partial_x^2 u + \partial_z u_z + \partial_z\log\overline{\rho}\left(u_z + \partial_x w\right) + \frac{1}{3}\left(\partial_x^2 u + \partial_x w_z\right)\right] \nonumber \\
&& \quad \quad \quad = - \pomega'\partial_x\Upsilon' - u\partial_x u - w u_z + \nu\left[ 2\partial_x u\partial_x\Upsilon' + \partial_x w \partial_z\Upsilon' + \partial_z\Upsilon' u_z - \frac{2}{3}\partial_x\Upsilon' \left(\partial_x u + w_z\right)\right], \label{PI horizontal dedalus} \\
&&\partial_t\Upsilon' + w\partial_z\log\overline{\rho} + \partial_x u + w_z = -u \partial_x\Upsilon' - w\partial_z\Upsilon', \label{PI continuity dedalus} \\
&&w\partial_z\log\overline{P} + \gamma\left(\partial_x u + w_z\right) - \frac{\chi}{C_V \overline{T}}\left(\partial_x^2 T' + \partial_z T'_z + T'_z\partial_z\log\overline{\rho}\right) \nonumber \\
&&\quad \quad \quad= \frac{\chi}{C_V \overline{T}}\left[\partial_x T' \partial_x\exp\Upsilon' + \left(\partial_x^2 T' + \partial_z T'_z + T'_z \partial_z\log\overline{\rho} \right)\left(\exp(\Upsilon')-1\right) + T'_z\partial_z\exp(\Upsilon')\right] \nonumber \\
&&\quad \quad \quad \quad+ \nu\frac{\exp(\Upsilon')}{C_V\overline{T}}\left[2\left(\partial_x u\right)^2 + \left(\partial_x w\right)^2 + u_z^2 + 2 w_z^2 + 2 u_z\partial_x w - \frac{2}{3}\left(\partial_x u + w_z\right)^2 \right], \label{PI constraint dedalus} \\
&&T' + \overline{T}\Upsilon' = \overline{T}\left(\exp(-\Upsilon') - 1 + \Upsilon' \right), \label{PI EoS dedalus} \\
&&T'_z - \partial_z T' = 0, \label{PI Tz dedalus}\\
&&w_z - \partial_z w = 0, \label{PI wz dedalus} \\
&&u_z - \partial_z u = 0. \label{PI uz dedalus}
\Eeq
Equations~\ref{PI vertical dedalus} \& \ref{PI horizontal dedalus} are the vertical and horizontal momentum equation respectively, equation~\ref{PI continuity dedalus} is the continuity equation, while equation~\ref{PI constraint dedalus} is the constraint equation.  Equation~\ref{PI EoS dedalus} is the fully nonlinear equation of state, while equations~\ref{PI Tz dedalus}--\ref{PI uz dedalus} define quantities which have second vertical derivatives within our first order system.  These are equivalent to equations~\ref{PI full momentum}--\ref{PI full constraint}.  Note that the momentum equations and the constraint equation have nonlinear diffusion terms which are treated explicitly.

For the PI equations with entropy diffusion, equations~\ref{PI constraint dedalus}--\ref{PI Tz dedalus} are replaced by
\Beq
&&w\partial_z\log\overline{P} + \gamma\left(\partial_x u + w_z\right) + \frac{\chi}{C_V\overline{T}}\left[ \overline{T}\partial_x^2\Upsilon' - \partial_z \tilde{Q}'_z - \tilde{Q}'_z\partial_z\log\overline{\rho}\right] \nonumber \\
&& \quad \quad \quad = -\frac{\chi}{C_V\overline{T}} \left[ \overline{T} \partial_x\Upsilon' \partial_x\exp(\Upsilon') +\left(\overline{T}\partial_x^2 -\partial_z\tilde{Q}'_z - \tilde{Q}'_z\partial_z\log\overline{\rho}\right)\left(\exp(\Upsilon')-1\right) - \tilde{Q}'_z\partial_z\exp(\Upsilon')\right] \nonumber \\
&&\quad \quad \quad \quad+ \nu\frac{\exp(\Upsilon')}{C_V\overline{T}}\left[2\left(\partial_x u\right)^2 + \left(\partial_x w\right)^2 + u_z^2 + 2 w_z^2 + 2 u_z\partial_x w - \frac{2}{3}\left(\partial_x u + w_z\right)^2 \right], \\
&& \tilde{Q}'_z + \overline{T}\partial_z \Upsilon' = 0.
\Eeq

The variables here have the same meaning as for the fully compressible equations.  The only additional variables are $\overline{\pomega}$ and $\pomega'$.  These are defined as follows.  Call the PI pressure perturbation $\pi'$, and recall that $\overline{\beta} = \overline{P}^{1/\gamma}$.  Define $\overline{\pi}$ by
\Beq
\overline{\beta}\grad\left(\frac{\overline{\pi}}{\overline{\beta}}\right) = \vec{g}\overline{\rho},
\Eeq
and define $\pi = \overline{\pi}+\pi'$.  Then $\pomega$ is defined as $\pomega=\pi/\rho$, and we can split it up as $\pomega=\overline{\pomega} + \pomega'$.  We have that $\overline{\pomega}$ satisfies the equation
\Beq
\overline{\beta}\grad\left(\frac{\overline{\pomega}}{\overline{\beta}}\right) + \overline{\pomega}\grad\log\overline{\rho} = \vec{g}.
\Eeq
In the PI equations, entropy is proportional to $\Upsilon'$, so the entropy boundary condition becomes $\Upsilon'=0$ on the top and bottom of the domain.  For the PI equations with temperature diffusion, we solve for the variables $u, u_z, w, w_z, \pomega', \Upsilon', T', T'_z$, and for the PI equations with entropy diffusion, we solve for the variables $u, u_z, w, w_z, \pomega', \Upsilon', \tilde{Q}'_z$.

\subsection{Anelastic Equations}

Our implementation of the AN equations with temperature diffusion is
\Beq
&& \partial_t w + \partial_z \pomega' - g \frac{S'}{C_P} - \nu \left[\partial_x^2 w + \partial_z w_z + 2 \partial_z\log\overline{\rho} w_z + \frac{1}{3} \left(\partial_x u_z + \partial_z w_z\right) - \frac{2}{3} \partial_z\log\overline{\rho} \left(\partial_x u + w_z\right) \right] \nonumber \\
&&\quad \quad \quad = -u\partial_x w - w w_z, \label{AN horizontal dedalus} \\
&& \partial_t u + \partial_x\pomega' - \nu \left[\partial_x^2 u + \partial_z u_z + \partial_z\log \overline{\rho} \left(u_z + \partial_x w\right) + \frac{1}{3}\left( \partial_x^2 u + \partial_x w_z\right)\right] = -u\partial_x u - w u_z, \label{AN vertical dedalus} \\
&&\partial_t S' + w \partial_z\overline{S} - \chi\left[ \partial_x^2\frac{S'}{C_P} + \frac{1}{C_P\overline{T}}\partial_x^2\pomega' - \frac{1}{\overline{T}}\partial_z \tilde{Q}'_z - \frac{1}{\overline{T}} \tilde{Q}'_z\partial_z\log\overline{\rho}\right] \nonumber \\
&&\quad \quad \quad = -u\partial_x S' - w\partial_z S' + \frac{\nu}{\overline{T}} \left[2\left(\partial_x u\right)^2 + \left(\partial_x w\right)^2 + u_z^2 + 2 w_z^2 + 2 u_z\partial_x w - \frac{2}{3}\left(\partial_x u + w_z\right)^2 \right], \label{AN entropy dedalus} \\
&&\tilde{Q}'_z + \overline{T} \partial_z\frac{S'}{C_P} + \frac{S'}{C_P}\partial_z\overline{T} + \frac{1}{C_P} \partial_z\pomega' = 0, \label{AN heat flux dedalus} \\
&&\partial_x u + w_z + w \partial_z\log\overline{\rho} = 0, \label{AN constraint dedalus} \\
&&w_z - \partial_z w = 0, \label{AN wz dedalus} \\
&&u_z - \partial_z u = 0. \label{AN uz dedalus}
\Eeq
Equations~\ref{AN vertical dedalus} \& \ref{AN horizontal dedalus} are the vertical and horizontal momentum equation respectively, equation~\ref{AN entropy dedalus} is the entropy equation.  Equation~\ref{AN heat flux dedalus} defines the heat flux, while equation~\ref{AN constraint dedalus} is the constraint equation.  Equations~\ref{AN wz dedalus} \& \ref{AN uz dedalus} define quantities which have second vertical derivatives within our first order system.

For the AN equations with entropy diffusion, equations~\ref{AN entropy dedalus} \& \ref{AN heat flux dedalus} are replaced by
\Beq
&&\partial_t S' + w \partial_z\overline{S} - \chi\left[ \partial_x^2\frac{S'}{C_P} - \frac{1}{\overline{T}}\partial_z \tilde{Q}'_z - \frac{1}{\overline{T}} \tilde{Q}'_z\partial_z\log\overline{\rho}\right] \nonumber \\
&&\quad \quad \quad = -u\partial_x S' - w\partial_z S' + \frac{\nu}{\overline{T}} \left[2\left(\partial_x u\right)^2 + \left(\partial_x w\right)^2 + u_z^2 + 2 w_z^2 + 2 u_z\partial_x w - \frac{2}{3}\left(\partial_x u + w_z\right)^2 \right], \\
&&\tilde{Q}'_z + \overline{T} \partial_z\frac{S'}{C_P} = 0.
\Eeq

The variables used here have all been defined for the FC \& PI equations.  One difference, however, is that $\pomega'$ is now a different pressure-type variable.  Calling the AN pressure perturbation $P'$, $\pomega'=P'/\overline{\rho}$.  This is different from the definition of $\pomega'$ used in the PI equations.  In both conduction models, we solve for the variables $u, u_z, w, w_z, \pomega', S', \tilde{Q}'_z$.

\bibliographystyle{apj}
\bibliography{TC}

\end{document}